\documentclass[a4paper,11pt]{article}
\usepackage{jheppub}

\usepackage{float}
\usepackage{subcaption}
\usepackage{amsmath}
\usepackage{color}
\usepackage{amssymb}
\usepackage{graphicx}
\usepackage{caption}
\usepackage{mathrsfs} 

\usepackage{hyperref}
\hypersetup{
  colorlinks = true, 
  urlcolor   = blue, 
  linkcolor  = blue, 
  citecolor  = blue  
}



\title{A Journey of Seeking Pressure and Forces in the Nucleon}

\author[a,b]{Xiangdong Ji}
\emailAdd{xji@umd.edu}

\author[a]{Chen Yang}
\emailAdd{cyang127@umd.edu}

\affiliation[a]{Department of Physics, University of Maryland, College Park, MD 20742, U.S.A.}
\affiliation[b]{T. D. Lee Institute, Shanghai 201210, China}

\abstract{
Momentum current density (MCD) $T^{ij}$ is a general physics concept 
describing the momentum conservation through momentum flow generated from both the kinetic motion of particles and the interacting forces among them. It has been suggested by M. Polyakov et al. that the MCD in the nucleon, characterized by the form factor $C/D$ of the QCD energy-momentum tensor, can be interpreted as the pressure and shear forces between adjacent parts of the system because the nucleon interior approximates a continuous medium. While intuitively appealing, we find that the interpretation is hard to justify from a detailed examination of the physical mechanisms for the momentum flow in QCD. After reviewing through a broad range of classical and quantum systems, we find that while thermal and/or quantum average of isotropic motion contributes to kinetic MCD a
pressure term proportional to $\delta^{ij}$, when there is an anisotropic motion, the pressure cannot simply be identified from the MCD tensor. Furthermore, kinetic pressure cannot be considered as the surface force between adjacent parts of a system. More importantly, at the scale of the nucleon dimension, the color forces among quarks and gluons is by no means short-ranged as in a continuous medium, and the resulting interaction MCD cannot be interpreted as normal or shear ``stress'' force, although an isotropic term from the QCD trace anomaly may be interpreted as a ``vacuum pressure.'' Following our previous study of force densities through divergences of kinetic MCDs, we affirm that the vacuum pressure term provides a confining potential on the quarks through color Lorentz forces. 
}

\begin{document}
\maketitle
\flushbottom

\section{Introduction}

The energy-momentum tensor (EMT) of quantum chromodynamics (QCD) has emerged as a powerful tool for probing the internal structure of the proton and neutron.  One of the earliest examples was the measurement of the spin-2 parts of quark and gluon EMT matrix elements through the deep-inelastic scattering~\cite{Georgi:1974wnj,Gross:1973zrg}. These matrix elements correspond to the longitudinal momentum fractions of the nucleon carried by quarks and gluons, respectively, in the infinite momentum frame (IMF)~\cite{Peskin:1995ev}. At the end of the 1980's, Jaffe and Manohar utilized the EMT to derive the spin sum rule of the nucleon in the IMF~\cite{Jaffe:1989jz}, a result later generalized by one of the present authors~\cite{Ji:1996ek}. The EMT has also played a central role in analyzing the nucleon’s mass structure~\cite{Ji:1994av,Ji:1995sv}. A crucial ingredient in these studies is the nucleon matrix elements of the QCD EMT in different momentum states, parametrized using the gravitational form factors (GFFs)~\cite{Kobzarev:1962wt,Pagels:1966zza,Ji:1996ek,Ji:1996nm}, which have become a focal point of both theoretical and experimental research efforts in recent years. While the early works about GFFs~\cite{Kobzarev:1962wt,Pagels:1966zza} concerned about the total EMT, whereas the recent QCD studies are mainly about the individual quark and gluon contributions to the bound state structure.

One of the important phenomenological advances has been the recognition that the spin-2 GFFs of the QCD EMT can be obtained from the generalized parton distributions (GPDs)~\cite{Ji:1996ek,Ji:1997gm,Muller:1994ses}. Moreover, it has been proposed to measure the GPDs through deeply-virtual Compton scattering~\cite{Ji:1996ek,Ji:1996nm} and diffractive meson production processes~\cite{Radyushkin:1996ru,Collins:1996fb} (see also~\cite{Ji:2016djn} for a historical overview and~\cite{Qiu:2022pla,Qiu:2022bpq} for references on the recent developments). There are also many lattice QCD calculations of GPDs and GFFs reported in the literature, including but not limited to~\cite{Mathur:1999uf,Hagler:2003jd,Gockeler:2003jfa,LHPC:2007blg,Shanahan:2018pib,Shanahan:2018nnv,Alexandrou:2020zbe,Pefkou:2021fni,Hackett:2023nkr,Hackett:2023rif,Lin:2023gxz}. 

Among these form factors, the $C$ form factor---also referred to as the $D$-term (denoted hereafter as $C/D$)---has gained special attention in recent years, with $D(0)$ dubbed as ``the last global unknown property''. A ``mechanical interpretation'' for this form factor was proposed in~\cite{Polyakov:2002wz,Polyakov:2002yz} and has sparked considerable interest~\cite{Ossmann:2004bp,Belitsky:2005qn,Goeke:2007fp,Goeke:2007fq,Cebulla:2007ei,Boffi:2007yc,Kim:2012ts,Mezrag:2013mya,Jung:2013bya,Son:2014sna,Jung:2014jja,dHose:2015eym,Perevalova:2016dln,Kumano:2017lhr,Hudson:2017xug,Hudson:2017oul,Kumano:2018xfe,Polyakov:2018guq,Kumano:2018jgo,Polyakov:2018aey,Panteleeva:2018ijz,Polyakov:2018zvc,Burkert:2018bqq,Lorce:2018egm,Shanahan:2018pib,Shanahan:2018nnv,Lorce:2018uyy,Anikin:2019kwi,Cosyn:2019aio,Polyakov:2019lbq,Detmold:2019ghl,Yanagihara:2019foh,Anikin:2019ufr,Azizi:2019ytx,Lorce:2019zyw,Mamo:2019mka,Lorce:2019ldg,Trinhammer:2019hoj,Neubelt:2019sou,Sun:2020wfo,Ozdem:2020ieh,Panteleeva:2020ejw,Rodini:2020pis,Alharazin:2020yjv,Varma:2020crx,Kim:2020nug,Fukushima:2020cmk,Zhang:2020ecj,Chakrabarti:2020kdc,Kim:2020lrs,Azizi:2020jog,Tong:2021ctu,Dutrieux:2021nlz,Zhang:2021mtn,Wang:2021dis,Roberts:2021nhw,Panteleeva:2021iip,Mamo:2021krl,Guo:2021ibg,Liu:2021gco,Burkert:2021ith,Freese:2021qtb,Metz:2021lqv,Kou:2021qdc,Gegelia:2021wnj,Zhang:2021tnr,Fiore:2021eav,Kim:2021jjf,Owa:2021hnj,Pefkou:2021fni,Kou:2021bzs,Epelbaum:2021ahi,Raya:2021zrz,Lorce:2021xku,Chavez:2021llq,Zhang:2021uak,Hashamipour:2021kes,Arrington:2021alx,More:2021stk,Liu:2021lke,Fu:2022rkn,Lorce:2022cle,Raya:2022eqa,Mamo:2022eui,Kim:2022syw,Fujita:2022jus,Varma:2022kbv,Wang:2022uch,Freese:2022ibw,Mezrag:2022pqk,Kim:2022wkc,Wang:2022ndz,Ji:2022exr,Won:2022cyy,Freese:2022fat,Panteleeva:2022uii,Tanaka:2022wzy,Alharazin:2022xvp,Shuryak:2023siq,Won:2023cyd,Liu:2023cse,More:2023pcy,Panteleeva:2023aiz,Chen:2023jbq,Burkert:2023wzr,Lorce:2023zzg,Moffat:2023svr,Choudhary:2023ihs,GarciaMartin-Caro:2023klo,Trinhammer:2023ykp,Pire:2023kng,Amor-Quiroz:2023rke,Panteleeva:2023evj,Kou:2023azd,Fu:2023ijy,More:2023vlb,Won:2023ial,Dehghan:2023ytx,Wang:2023fmx,Cao:2023ohj,Hackett:2023nkr,Hackett:2023rif,Burkert:2023atx,He:2023ogg,Singh:2023hgu,Wang:2023bjp,Xu:2023izo,Alharazin:2023uhr,GarciaMartin-Caro:2023toa,Chen:2024adg,Raya:2024ejx,Nair:2024fit,Xu:2024hfx,Diehl:2024bmd,Liu:2024vkj,Cao:2024fto,Hu:2024edc,Yao:2024ixu,Dutrieux:2024bgc,Wang:2024abv,Krutov:2024adh,Son:2024uet,Broniowski:2024mpw,Freese:2024rkr,Klest:2025rek,Lorce:2025oot,Goharipour:2025lep,Deng:2025fpq,Dehghan:2025ncw,Broniowski:2025ctl,Fujii:2025aip,Sain:2025kup,Sugimoto:2025btn,Choi:2025rto,Freese:2025tqd,Fujii:2025tpk,Tanaka:2025pny,Fujii:2025paw}. This interpretation employs an analogy to fluid dynamics, assigning the trace and traceless parts of the QCD momentum current density (MCD) $T^{ij}(\vec{r})$ with the physics of ``pressure'' and ``shear force'' respectively. Furthermore, this perspective has also been applied to the hadron mass decomposition, leading to two-term decompositions~\cite{Lorce:2017xzd,Lorce:2018uyy,Lorce:2019ldg,Lorce:2021xku} and other frameworks~\cite{Cosyn:2019aio,Rodini:2020pis,Liu:2021gco}, which differ from the original four-term decomposition~\cite{Ji:1995sv,Ji:1994av} by combining the contributions of different physical origins. A ``mechanical stability condition'' was also proposed, suggesting that the $D$-term form factor should be negative, although recent investigations of the hydrogen atom~\cite{Ji:2022exr,Czarnecki:2023yqd}, spin-$3/2$ particles such as the $\Delta$ resonance and $\Omega^-$~\cite{Fu:2022rkn,Fu:2023ijy,Wang:2023bjp} have reported $D$-terms of the opposite sign. 

Despite the obvious attraction of the ``mechanical interpretation'', we find the identification of pressure and forces directly in terms of MCD $T^{ij}$ to be conceptually puzzling. Some basic questions come to mind: 
\begin{itemize}
    \item While MCD is a general concept for any physical systems, not all MCDs can be interpreted as pressure and shear forces, as in the examples of the static electric field of a point charge and a single moving particle, where no pressure is obviously present. Under what conditions then, can an MCD be interpreted as pressure or shear forces between adjacent parts of a system? 
    \item Why invoke a classical mechanical stability criterion for a quantum system, especially when classical mechanics was abandoned because of its inadequacy at sub-atomic scales? If the proposed stability condition depends on the interaction range~\cite{Perevalova:2016dln}, how should one understand the stability of systems with different ranges of forces? 
    \item The EMT is not unique for any physical systems and most discussions have focused on the one that is supposed to couple to gravity. Does the pressure and force interpretation apply only to this particular EMT, and if so, why? 
\end{itemize}
While some of these questions have been discussed in a recent paper~\cite{Lorce:2025oot}, we find the relevant answers not yet fully satisfactory. 

To address the above questions in depth, we begin by examining the physics of the momentum continuity equation, a general consequence of translational symmetry of physical space. It is well known that for any isolated physical system, its total momentum $\vec{K}$ is a constant of motion. By introducing the momentum density $k^i = T^{0i}(\vec{r},t)$ and the MCD $T^{ij}(\vec{r},t)$ ($i,j=1,2,3$ are spatial indices, representing the $j$-th component of the momentum flowing in the $i$-th direction), this conservation law can be expressed in a local differential form as, 
\begin{equation}
    \frac{\partial T^{0j}}{\partial t} + \nabla_i T^{ij} = 0 \ ,\label{eq:Continuity}
\end{equation}
where $\nabla_i\equiv \partial_i$ is the spatial differential operator. The above equation depicts a dynamical picture of local momentum conservation, providing a window into the structure and dynamics of the system. The above equation directly leads to a global result for any isolated static system,
\begin{equation} 
    \int T^{ii}(r) {\rm d}^3 \vec{r} = 0 \label{eq:von-Laue} \ ,
\end{equation}
i.e., the 3-trace of the total MCD $T^{ii}(\vec{r})$ must overall be equally positive and negative throughout space. It has been called the von-Laue stability condition~\cite{Laue:1911lrk}, discussed, e.g., in the stability of the Skyrme model~\cite{Cebulla:2007ei}. This equation has also been considered in the context of the virial  theorem~\cite{Lorce:2021xku}, which was early referred to as a relation between scalar and tensor contributions to mass in quantum field theories (QFTs)~\cite{Ji:1994av}. 

We examine the physics of MCDs in various classical and quantum systems, and focus particularly on the possible identification of certain parts as pressure and mechanical forces between adjacent parts of a system. The momentum flow in general originates from the kinetic motion of particles and the interacting forces among them. 
Contributions to the kinetic part $T^{ij}_K$ come from both isotropic and anisotropic motion. The thermal and/or quantum mechanical average of the isotropic motions usually generates a diagonal MCD term proportional to $\delta^{ij}$, identified as the kinetic pressure $p_K$. However, an anisotropic motion as in the velocity field of a fluid contributes a kinetic MCD that cannot easily be interpreted as pressure, as a laser beam has a momentum flow but zero pressure. In general, identifying a pressure term in a kinetic MCD with anisotropic motion is not straightforward, and one has to examine the physics details. In the example of the laser beam, one cannot simply take the monopole term $T^{ii}/3$ as its pressure. 

The kinetic pressure cannot be interpreted as a surface force between adjacent parts of a system. The simplest example is an ideal gas system in which there is a pressure but there are no interactions among particles. 

The interaction MCDs $T^{ij}_{I}$ play a role analogous to a potential $V(\vec{r})$ of a conservative force $\vec{F} =-\nabla V(\vec{r})$, and are not uniquely defined in general. Independent of the range of the potentials, if an interaction MCD has only a diagonal term proportional to $\delta^{ij}$, it has been identified as an interaction pressure $p_I$. The total pressure is then $p=p_K+p_I$ which will generate a surface force on the boundary of a box or a pressure gauge due to the discontinuity of the momentum flow. However, for an interaction with a range comparable to the size of a system, neither the pressure $p_I$ nor the interaction MCD tensor can be regarded as a surface force between adjacent parts of a system.  The interaction MCD has no other mechanical significance other than that it describes the momentum flow. Only when the range of the potential is negligible compared with the size of a system, such as in a continuous medium with contact interactions, the interaction MCD then has a physical meaning as a contact surface force. 

With these findings, we analyze separate contributions to the symmetric and gauge-invariant QCD MCD in the nucleon and study their physical significance. The quark MCD reveals an anisotropic transport of momentum by quark motions. The gluon MCD includes contributions from the momentum flow generated by radiative gluons, static interactions, as well as the trace anomaly. Among these, the radiative part arises from the anisotropic kinetic transport through physical gluons. In contrast, the static interaction part comes from the Coulomb-like, long-range color forces within the nucleon---where ``long-range'' refers to a scale comparable to the nucleon radius. Unfortunately, there is no frame-independent and gauge-invariant separation of these two parts, and therefore we combine them together as the tensor gluon MCD. The QCD trace anomaly contributes an isotropic term in MCD, which can be interpreted as a vacuum pressure, arising from the change of the QCD vacuum in the presence of the valence quarks~\cite{Ji:1995sv,Liu:2021gco}. 

Therefore, we find no evidence that any part of QCD MCD or the total has a direct interpretation as mechanical pressure or shear surface forces. The kinetic contributions of quarks and radiative gluons are anisotropic, and there is no physical ground for their monopole parts to be identified as a pressure inside the nucleon. The strong force within the nucleon, being long-range, cannot be reduced to contact forces relevant for 2D surface forces as in a continuum medium. The anomaly contribution can be regarded as a pressure, but functions as an external potential since there is no sharp boundary as in the M.I.T. bag model. 
Finally, physical forces must be identified through the divergences of the MCDs, which
are balanced (summed to zero) everywhere inside the nucleon. The color-Lorentz force densities acting on the quarks, which have been reported in our previous work~\cite{Ji:2025gsq}, will be reconsidered again for non-vanishing light-quark masses. 

The presentation of the paper is as follows. In Sec.~\ref{sec:MCD}, we provide a general discussion of the MCD, its relation to forces and intrinsic ambiguities, with a focus on the particular formulation believed to couple to gravity. In Sec.~\ref{sec:MCD-Classical}, we review the physics of MCDs in classical systems, including matter systems of gases, solids and liquids, as well as electromagnetic fields, demonstrating that the MCD receives contributions from both particle kinetic motion and interacting fields. In Sec.~\ref{sec:MCD-Quantum}, we extend our discussions to quantum systems, including both single-particle and many-body cases. In particular, we show that quantum wavefunctions for single particles display ordered flows, invalidating the concepts of pressure-volume work. We also consider the particularly interesting example of quantum Bose liquids. In Sec.~\ref{sec:MCD-QCD}, we study the QCD MCD in the proton. We examine the confining color-Lorentz force acting on quarks, in which the trace anomaly plays a dominant role. In Sec.~\ref{sec:Argument}, we examine recent proposals concerning the mechanical interpretation of the MCD and present our critical assessment. We summarize the paper in Sec.~\ref{sec:Conclusion}.

\section{Momentum Current Density: Relation to Forces and Potentials} \label{sec:MCD}

In this section, we discuss a range of topics concerning the definition and physics of MCDs, with the aim of understanding the underlying mechanisms of the momentum flow. In general, MCDs receive contributions from two distinct sources: the kinetic motion of particles and the interactions or forces among them. While the former can be unambiguously defined, the interaction MCDs inherently involve a degree of arbitrariness related to, e.g., boundary conditions. Only one specific formulation---namely, the one that couples to gravity---is considered physically ``measurable''. This already suggests that the total MCD of a system may not possess a uniquely-meaningful mechanical interpretation. 

We begin with the local momentum conservation in the non-relativistic limit, where the momentum density of a system is entirely carried by matter particles. We present a bottom-up method to establish constraint equations governing the interaction MCD of mechanical forces, using Newton's second law and the momentum continuity equation. However, these equations are not sufficient to uniquely determine all components of the MCD tensor, leaving an infinite number of possible MCDs differing by what may be viewed as analogous to ``gauge'' degrees of freedom. A specific form of the MCD can only be obtained by specifying a ``gauge condition''.

For relativistic systems, it is more natural to adopt the field-theoretic framework based on Lagrangian densities~\cite{Peskin:1995ev}. According to Noether's theorem, three-momentum conservation arises from the spatial translational symmetry of the action. The associated conserved EMTs, and their spatial components identified as MCDs, can be systematically derived from this symmetry. However, it is well known that Noether's theorem does not yield a unique conserved current, as one may freely add superpotential terms to the current without violating the conservation law~\cite{BELINFANTE1940449,Blaschke:2016ohs}.

On the other hand, Einstein's general theory of relativity singles out a particular version of EMT (also called the metric EMT in the literature) as the physical source for gravitational fields, responsible for generating curved spacetime. For theories of interest here, such as quantum electrodynamics (QED) and quantum chromodynamics (QCD), it has been shown that the metric EMT coincides with the Belinfante-Rosenfeld improved EMT derived from the canonical form~\cite{Blaschke:2016ohs}. While it is difficult to argue whether this ``chosen'' MCD holds any special mechanical significance for the system itself, the individual contributions do carry established physical interpretations, which we will explore further in Sec.~\ref{sec:MCD-QCD}. 

\subsection{Non-Relativistic MCD} \label{subsec:NRMCD}

We begin our discussion of MCD with non-relativistic isolated systems made of material particles. For such a system with $N$ free ones, one can directly write down the momentum density and the kinetic MCD as~\cite{Weinberg:2021kzu}, 
\begin{align}
    T^{0i}_{K} (\vec{r},t)& = \sum_a k_a^i(t) \delta^{(3)} \left( \vec{r} - \vec{r}_a(t) \right)  \ , \label{eq:MD-Particle} \\
    T^{ij}_{K} (\vec{r},t) & = \sum_a v_a^i(t) k^j_a(t) \delta^{(3)} \left( \vec{r} - \vec{r}_a(t) \right) \ , \label{eq:MCD-Particle}
\end{align}
where $k_a^i = m_a v_a^i$ $(a=1,2,...,N)$ and $m_a,\vec{r}_a,\vec{v}_a,\vec{k}_a$ are the mass, position, velocity and momentum of the $a$-th particle, respectively. Clearly, $T^{ij}_{K}$ is symmetric in indices $i$ and $j$, manifestly describing a flow of $j$-th component of the momentum along the $i$-th direction, through the kinetic motion of particles. One can easily check that the momentum continuity equation in Eq.~(\ref{eq:Continuity}) is trivially satisfied. 

When interactions are present between particles---such as through a two-body potential $V_{ab}$---the global momentum of the system remains conserved; however, the naive form of the local momentum continuity equation is no longer valid, 
\begin{equation}
 \frac{\partial T^{0j}_{K}(\vec{r},t)}{\partial t} + \nabla_i T^{ij}_{K}(\vec{r},t) = \sum_a F^j_a(t) \delta^{(3)} \left(\vec{r}-\vec{r}_a(t)\right) \equiv {\cal F}^j(\vec{r})\ , \label{eq:Force}
\end{equation}
where we have used Newton's second law and $F_a^j$ is the internal force exerted on the $a$-th particle by the others. This equation yields the internal {\it force density} ${\cal F}^j(\vec{r})$ acting within the system of particles. It illustrates that momentum is continuously transferred among particles through interactions (or through underlying fields as in the case of electromagnetism). As a result, local conservation holds only when these transfer effects are properly accounted for. To restore the local form of the continuity equation, one may introduce an interaction MCD $T^{ij}_{I}$ such that,
\begin{equation}
    \frac{\partial T^{0j}_{K}(\vec{r},t)}{\partial t} + \nabla_i \left(T^{ij}_{K}(\vec{r},t) + T^{ij}_{I}(\vec{r},t)\right) = 0 \ , \label{eq:Continuity-NR}
\end{equation}
where the new tensor satisfies the following three constraint equations, 
\begin{equation}
    \nabla_i T^{ij}_{I} =  - {\cal F}^j(\vec{r}) \ . \label{eq:MCD-Int}
\end{equation}
By integrating Eq.~(\ref{eq:Continuity-NR}) over an arbitrary spatial volume, one finds that the change in momentum within this volume is related to the total momentum current flowing out of the surface, either via the motion of particles or through interacting forces. 

The above equation suggests that the interaction MCD functions analogously to a potential, in the sense that only its divergence is physically meaningful through its relation to the force density. It is well-known that gravitational and electric potentials lack absolute meaning, as adding an arbitrary constant does not generate observable consequences. In the present case, even greater freedom exists: the three momentum continuity equations cannot determine $T^{ij}_{I}$ since it has nine degrees of freedom. Even under the assumption that $T^{ij}_{I}$ is symmetric in $i$ and $j$, there remain six independent components. For any given solution $T^{ij}_{I}$, one is free to add an additional term~\cite{landau2013fluid},
\begin{equation}
    T^{\prime ij}_{I} = T^{ij}_{I} + \partial_k \chi^{[ki]j} \ ,
\end{equation}
where $\chi^{[ki]j} = \chi^{kij} - \chi^{ikj}$ has been called the superpotential in the literature~\cite{Blaschke:2016ohs}. Clearly, this new MCD is also a valid solution of the momentum continuity equation and yields the same momentum conservation, albeit with a different local structure. One can verify that $\partial_k\chi^{[ki]j}$ has six independent components, which, when combined with the above three constraint equations, can fully determine all components of $T^{ij}_{I}$. Therefore, without additional input, there is no compelling reason to prefer one form of local conservation law over the other. 

Additional inputs---such as the boundary conditions and mechanical models for the force density---may be considered as a ``gauge choice''. In the following sections, we will explicitly present examples showing that the same force density can give rise to entirely different interaction MCDs, depending on the selected force models and boundary conditions. Only one of them admits an interpretation as surface forces---when the system can be described as a static continuous medium with short-range or contact interactions.

\subsection{Relativistic MCD and Coupling to Gravity}

For relativistic systems, Newton's law and forces are not the most natural concepts. Instead, theories such as the standard model of particle physics are formulated within the framework of QFT, which employs fields and Lagrangian densities. In this context, momentum conservation arises as a consequence of the spacetime translational symmetry of the action. The corresponding local conservation follows from the well-known Noether's theorem~\cite{noether1971invariant}. 


While Noether's theorem determines uniquely the total conserved charges,  the conserved current densities can differ by a divergence-free term, called the superpotential. For example, one can define a new conserved EMT as, 
\begin{equation}
    T^{\prime \mu\nu} = T^{\mu\nu} + \partial_\rho \chi^{[\rho\mu]\nu}[\Phi] \ ,
\end{equation}
where $\chi^{[\rho\mu]\nu}$ is the superpotential, a function of fields $\Phi$, anti-symmetric in $\rho,\mu$. We may call the ambiguities in choosing the superpotential as ``gauge degrees of freedom''. For example, in gauge theories like QED and QCD, the canonical EMT is neither gauge-invariant nor symmetric. Such a problem is remedied by using the Belinfante-Rosenfeld improvement~\cite{BELINFANTE1940449}, where one constructs a superpotential from the spin angular momentum tensor $S^{\rho\mu\nu}$, 
\begin{equation}
    \chi^{\rho\mu\nu} = \frac{1}{2}\left(S^{\rho\mu\nu}-S^{\mu\rho\nu}-S^{\nu\rho\mu}\right) \ ,
\end{equation}
where $S^{\rho\mu\nu}$ is anti-symmetric in $\mu,\nu$. However, there are still remaining ambiguities in choosing the superpotential while maintaining its symmetric and gauge-invariant property~\cite{Blaschke:2016ohs}. One certain choice is $\chi^{\rho\mu\nu} = \left(g^{\nu\rho}\partial^{\mu}-g^{\nu\mu}\partial^{\rho}\right)F[\Phi]$, which gives an additional term as follows, 
\begin{equation}
    T^{\prime \mu\nu} = T^{\mu\nu} +  \left( \partial^\mu \partial^\nu - g^{\mu\nu}\partial^2 \right)F[\Phi] \ , \label{eq:extra-EMT}
\end{equation}
where $F[\Phi]$ is a gauge-invariant functional of fields. These additional degrees of freedom correspond to moment series in multipole expansion~\cite{Ji:2021mfb}. 

A superpotential term may also arise from the total derivative term in the Lagrangian $\mathcal{L}' = \mathcal{L} + \partial_\mu S^\mu[\Phi]$, where $S^\mu[\Phi]$ is a functional of fields. It can be directly shown that the equations of motion remain unchanged under this modification; see Appendix.~\ref{sec-app:EoM} for a detailed derivation. However, this total derivative does contribute to the canonical conserved current. Specifically, according to Noether's theorem, there is an additional contribution to the EMT~\cite{Gieres:2022cpn}, 
\begin{equation}
    \delta T^{\mu\nu} = \partial^{\nu}S^\mu - g^{\mu\nu}\left( \partial_\rho S^\rho \right) \equiv \partial_\rho \chi^{\rho\mu\nu} \ ,
\end{equation}
where $\chi^{\rho\mu\nu}=S^\mu g^{\rho\nu}-S^{\rho}g^{\mu\nu}$ is the corresponding superpotential. For example, suppose one has $S^\mu = \partial^\mu F[\Phi]$ and $F[\Phi]$ is gauge invariant, the extra EMT in Eq.~(\ref{eq:extra-EMT}) is reproduced. 



There is a particular version of EMT that serves as the source of gravity. In general relativity, this EMT appears in the Einstein equation, 
\begin{equation}
    G_{\mu\nu} + \Lambda g_{\mu\nu} = 8\pi G T^{\mu\nu} \ ,
\end{equation}
where $G_{\mu\nu}$ is the Einstein tensor, $\Lambda$ is the cosmological constant, $g_{\mu\nu}$ is the metric tensor, $G$ is the Newtonian constant of gravitation and $T^{\mu\nu}$ is the Einstein–Hilbert (metric) EMT. For a field theory in curved spacetime with action $S_M$, such an EMT can be derived through, 
\begin{equation}
    T^{\mu\nu}[\psi,g] \equiv \frac{2}{\sqrt{-g}} \frac{\delta S_M[\psi,g]}{\delta g_{\mu\nu}} \ .
\end{equation}
This is similar to the gauge principle, wherein a conserved current arises when promoting a global symmetry to a local one. In certain simple field theories involving only first-order derivatives---such as QED, QCD or scalar fields---the above metric EMT is the same as the Belinfante-Rosenfeld improved EMT. 

However, as far as the momentum conservation of a system is concerned, the gravity-coupled EMT does not appear to have any special status, aside from the fact that it could, in principle, be measured through gravitational experiments. On the other hand, the MCD that admits an interpretation in terms of surface forces---as in the case of liquids and solids---may not be the one that couples to gravity.


\section{MCDs and Pressure/Forces in Classical Systems}\label{sec:MCD-Classical}

In this section, we examine the components of MCDs in classical systems, including matter systems such as gases, liquids and solids, as well as electromagnetic and other classical field theories. Our goal is to identify the physics of momentum flow by analyzing the contributions from both kinetic motion of constituent matter particles and the interactions among them. 

A basic feature of matter systems is their enormous number of degrees of freedom, originating from the individual atoms or molecules comprising them. A commonly adopted approach in analyzing solids and liquids---and, to some extent, gases---relies on the assumption of {\it scale separation}. This assumption permits the definition of a continuous distribution of volume elements (referred to hereafter as ``elements'') that are significantly smaller compared to the macroscopic scale of the system, yet large enough to encompass a statistically meaningful number of constituent particles. Consequently, each element represents a sub-thermodynamic system in {\it local equilibrium}, described by the position-$\vec{r}$ dependent intensive observables such as density ($\rho$), velocity ($\vec{v}$), temperature ($T$) and pressure ($p$), or more generally, the stress tensor ($\sigma^{ij}$)~\cite{landau2013fluid}. A key feature of continuous-medium models is that forces between elements are zero-range or occur only through contact, leading to simple mechanical interpretations of the resulting {\it phenomenological} MCDs. Our study finds that,
\begin{itemize}
    \item MCDs for gases and liquids have an isotropic piece $\delta^{ij}p$ as pressure from both microscopic motion of particles and interactions among them. However, there is an anisotropic contribution from macroscopic (kinetic) motion of elements described by the velocity field $\vec{v}(\vec{r})$ which cannot be regarded as a pressure. In particular, when a macroscopic motion is present, the pressure cannot be identified as a monopole term in the total MCD tensor. The pressure is in general not a pure surface force among difference parts of a system, particularly the kinetic one in the case of gases. 
    \item In liquids, the macroscopic motion can also generate a stress tensor (surface/shear forces) as a part of MCDs due to momentum flow exchanges between macroscopic and microscopic degrees of freedom through a dissipative mechanism. In solids, the stress tensor as a part of MCD due to elastic deformation is indeed a surface force. In both cases, short-range contact interaction is the key for the physical interpretation of the interaction MCD, whereas shear forces in liquids require the scale separation of degrees of freedoms. 
\end{itemize}

On the other hand, the momentum flow in classical field theories has a completely different character. The definition of momentum is now associated with dynamics
of fields as in $\pi(x)=\partial {\cal L}/\partial \dot{\phi}(x)$, and the flow of momentum is associated with transport of field disturbances. Therefore, the MCD of a classical field is physically similar to the 
macroscopic motion term in gases and liquids, which can be made manifestly through quantization. The concepts of pressure and forces are not directly applicable unless there are local nonlinear interactions. The static forces of electromagnetism and gravity are associated with non-propagating classical fields which contribute to the MCD under a particular choice of boundary conditions, namely the explicit presence of static electric and gravitational fields. This contribution, historically called stress tensor, has no real significance as pressure or forces as aether does not physically exist. A simple example is a bound state of two opposite electric charges revolving around each other in a planetary motion where the static electric field does not generate pressure or surface forces on any part of the system. Their physical significance as ``stress tensor'' is simply to generate forces through the divergence of the corresponding MCD~\footnote{Therefore, the self-field contribution to MCD is completely unobservable.}, similar to the external gravitational potential acting on a system. 

\subsection{Momentum Flow in Gases: Kinetic and Interaction Pressures} 

Gases consist of a collection of molecules in a container of volume $\mathcal{V}$, with little interaction between them. As a result, each molecule can move freely throughout the container without restrictions. In the case of an ideal gas, there are no intermolecular interactions or collisions at all; the only force exerted on gas particles arises from collisions with the container walls. Consequently, the momentum flow within the gas occurs exclusively through the kinetic motion of particles carrying the momentum densities $T^{0i}_K$ and the non-relativistic kinetic MCD $T^{ij}_{K}$ as in Eqs. (\ref{eq:MD-Particle}) and (\ref{eq:MCD-Particle}), respectively.  

For an ideal gas in local equilibrium, the velocity of each molecule $\vec{v}_a$ can be written as a superposition of macroscopic drift $\vec{v}(\vec{r})$ and microscopic random motion $\delta \vec{v}_a$. The statistical average over microscopic degrees of freedom gives, 
\begin{align}
     \langle T^{ij}_{K} \rangle (\vec{r})
     = \rho(\vec{r}) v^i(\vec{r}) v^j(\vec{r}) +  p_K(\vec{r})\delta^{ij} \ , \label{eq:MCD-idealgas}
\end{align}
where $\rho(\vec{r})$, $n(\vec{r})$, $\varepsilon_K(\vec{r})$ and $p_K(\vec{r})=\frac{2}{3} n(\vec{r}) \varepsilon_K (\vec{r})$ are the local mass, number, average kinetic energy densities of particles, and kinetic pressure, respectively. While the first term represents the anisotropic macroscopic transport of particles, the second term arises from the isotropic microscopic thermal motion. It is quite clear that one cannot obtain a pressure by isolating the monopole term $T^{ii}_K/3$ when $\vec{v}\ne 0$. The dynamical pressure, often introduced to describe the effect of the moving field, is $\rho v^2/2$, is not the same~\cite{landau2013fluid}. 

\begin{figure}[ht]
    \centering
    \begin{subfigure}[b]{0.35\textwidth}
        \centering
        \includegraphics[width=0.58\linewidth]{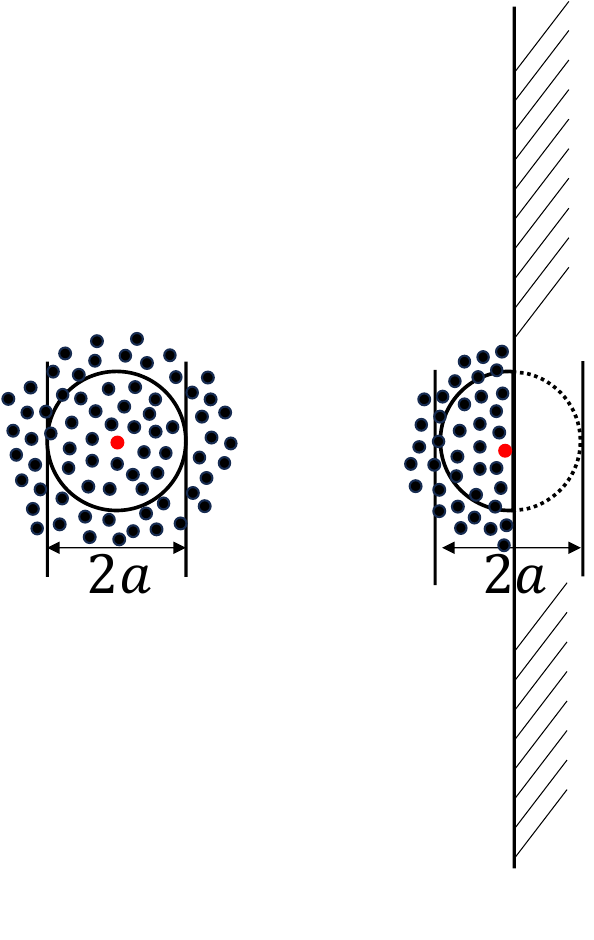}
        \caption{}
    \end{subfigure}
    \begin{subfigure}[b]{0.35\textwidth}
        \centering
        \includegraphics[width=1\linewidth]{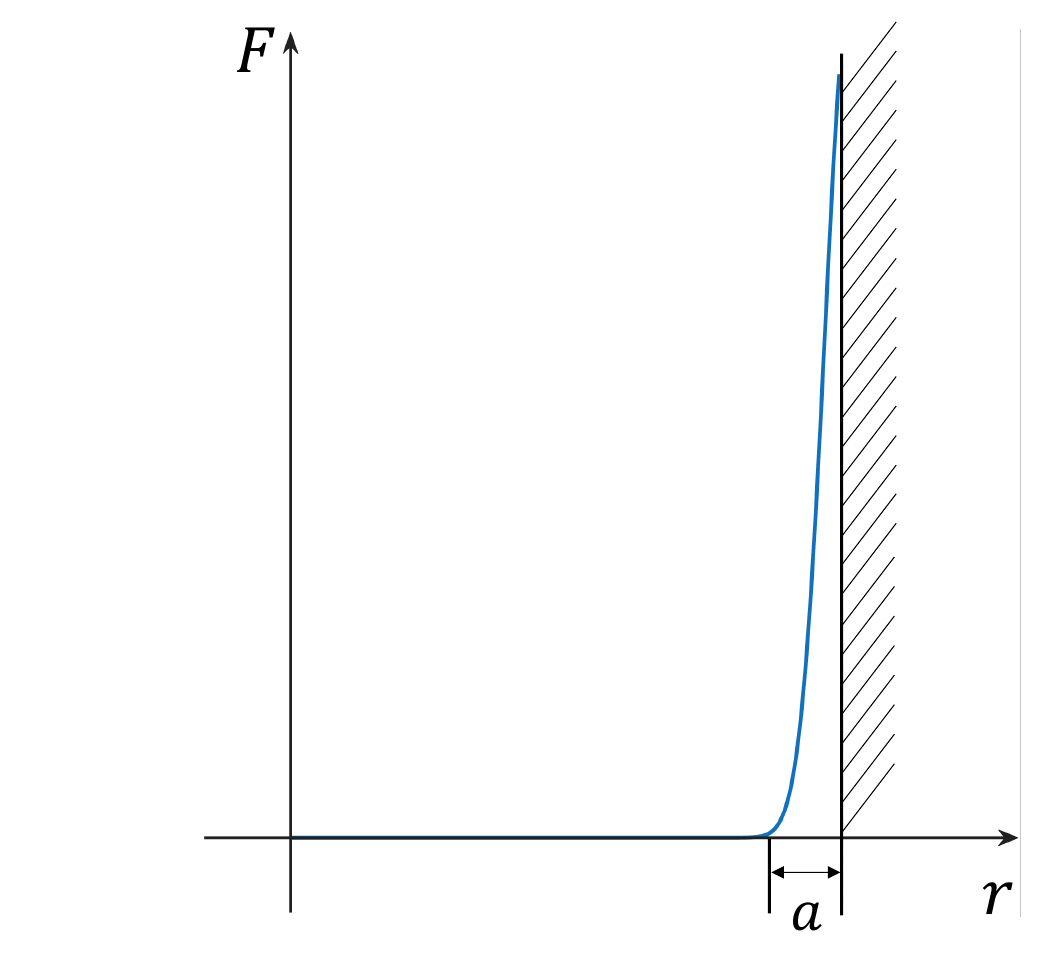}
        \caption{}
    \end{subfigure}
    \caption{Figure (a) shows the force on a molecule (red dot) located at different positions within the container, from other molecules in a sphere of radius equal to the interaction range $a$. Figure (b) shows that the force density is non-vanishing only near the container boundary. The force range $a$ is effectively taken as zero at macroscopic level.}
    \label{fig:gas-Interaction}
\end{figure}

When interactions among particles are introduced, an additional interaction MCD is needed to fulfill the continuity equation within the gas. This interaction MCD can be directly solved through Eq.~(\ref{eq:MCD-Int}), 
\begin{equation}
    \nabla_{i}T_{I}^{ij}=-\mathcal{F}^{j}(\vec{r})=\sum_{a}\left[\sum_{b\neq a}\nabla_{a}^{j}V\left(|\vec{r}_{a}-\vec{r}_{b}|\right)\right]\delta^{(3)}(\vec{r}-\vec{r}_{a})
\end{equation}
where $V$ is the two-body potential among gas particles. The above equation gives the internal force density acting on particles at position $\vec{r}$. Relative to the size of the system, the range of molecular forces is effectively zero. As a result, the force density is essentially zero inside the gas, becoming non-vanishing only near boundaries within the force range, where the interaction on a particle becomes asymmetrical, as shown in FIG.~\ref{fig:gas-Interaction}. 

By performing a statistical average over particles in different positions and neglecting higher-order correlations, one has, 
\begin{align}
    \nabla_{i}T_{I}^{ij}
    &= \sum_{a}\sum_{b\neq a} \int_{\mathcal{V}}\frac{{\rm d}^{3}\vec{r}_{a}}{\mathcal{V}}\int_\mathcal{V}\frac{{\rm d}^{3}\vec{r}_{b}}{\mathcal{V}}\left[\nabla_{a}^{j}V(|\vec{r}_{a}-\vec{r}_{b}|)\right]\delta^{(3)}(\vec{r}-\vec{r}_{a})e^{-V(\vec{r}_{a}-\vec{r}_{b})/kT}\\
    &= -\frac{N^{2}kT}{2 \mathcal{V}^{2}}\int{\rm d}^{3}\vec{x}\left[e^{-V(\vec{x})/kT}-1\right] \nabla^j\theta_{\cal V} \equiv p_I \nabla^j\theta_{\cal V}
\end{align}
where $\theta_{\cal V}$ is 1 inside and 0 outside the volume $\mathcal{V}$. The last factor in the first line comes from the Boltzmann distribution and the constant $-1$ is added to the integral to constrain the infrared behavior at large distances. Notably, since the interaction potential is effectively zero-range at macroscopic level, the statistical average yields non-zero result only when $\vec{r}$ is exactly on the boundary, which gives the factor $\nabla^j\theta_{\cal V}$ in this case. On the other hand, there is also one particular solution to Eq.~(\ref{eq:MCD-Int}), which leads to the interaction MCD for gases after the statistical average as,
\begin{equation}
    T_{I}^{ij}(\vec{r})=\sum_{a}\nabla^{i}\left[-\frac{\nabla_{a}^{j}V(\vec{r}_{a})}{4\pi|\vec{r}-\vec{r}_{a}|}\right]=p_{I}\theta_{\cal V}\delta^{ij}+\left(\nabla^{2}\delta^{ij}-\nabla^i \nabla^j\right)\int_{\cal V}{\rm d}^{3}\vec{x}\frac{p_{I}}{4\pi|\vec{r}-\vec{x}|}
\end{equation}
Therefore, the solutions for the interaction MCD can in general be written as,
\begin{equation}
    \langle T_{I}^{ij} \rangle \equiv p_I \theta_{\cal V} \delta^{ij} + \partial_k \chi^{[ki]j}_{I}\ ,
\end{equation}
where $\chi^{[ki]j}_I$ is the undetermined superpotential. The interaction pressure $p_I$ is exactly the second virial coefficient~\cite{pathria2017statistical}.

We further explore the meaning of the kinetic and interaction MCDs by focusing on a homogeneous ideal gas in a container,
\begin{equation}
    \left\langle T_{{\rm gas}}^{ij}\right\rangle \equiv \langle T_{{K}}^{ij}+T_{I}^{ij} \rangle = \delta^{ij}\left(p_K+p_I\right)\theta_{\cal V} + \partial_k \chi^{[ki]j}_{I}\ ,
\end{equation}
which is {\it not conserved} at the container walls or at the tip of a piezometer inserted within. This non-conservation arises from the {\it elastic collisions} or {\it contact interactions} between the gas particles and the boundary, which result in momentum transfer between them. To account for these collisions, one may further introduce an external interaction MCD, denoted as $T^{ij}_{\rm ext}$. The total MCD, $T^{ij}_{\rm tot}= T^{ij}_{\rm gas}+T_{\rm ext}^{ij}$, is then conserved throughout all of space. Newton's third law yields, 
\begin{equation} 
    \partial_i T^{ij}_{\rm gas} = -\partial_i T^{ij}_{\rm ext} = -\left( p_K + p_I \right) n^j \delta_S\ ,\label{eq:div-MCD-Gas}
\end{equation}
where $n^j$ is the normal vector of the surface pointing outward from the container, $\delta_S=-\hat n\cdot\partial \theta_{\cal V}$ is the Dirac delta function on the boundary. Similarly, one solution to this boundary MCD with ambiguities from the superpotential is given by,
\begin{equation}
    T^{ij}_{\rm ext} = - \left( p_K+ p_I \right) \delta^{ij}  \theta_{\cal V} + \partial_k \chi^{[ki]j}_{\rm ext} \ . \label{eq:MCD-Gas-Boundary}
\end{equation}
The total conserved MCD of the gas-container system is now, 
\begin{equation}
    T^{ij}_{\rm tot}= \partial_k (\chi^{[ki]j}_I + \chi^{[ki]j}_{\rm ext})  \ , 
\end{equation}
which receives contributions purely from the superpotential. One simple choice is,
\begin{equation}
    \chi^{[ki]j}_{I}=\chi^{[ki]j}_{\rm ext}=0, ~~~T^{ij}_{\rm tot}= T^{ij}_K + T^{ij}_I + T^{ij}_{\rm ext} = 0 \ ,
\end{equation}
and the von Laue condition in Eq.~(\ref{eq:von-Laue}) is trivially satisfied. The boundary contributes a negative MCD (Eq.~(\ref{eq:MCD-Gas-Boundary})) within the gas, which should not be interpreted as a ``negative pressure'', but rather as a form of external MCD potential. A similar situation will be encountered in the M.I.T. bag model, as discussed in the following section. 

As indicated in Eq.~(\ref{eq:Force}), the divergence in Eq.~(\ref{eq:div-MCD-Gas}) gives the external force density acting on gases. By Newton's third law, this force density corresponds to the reaction force exerted on the wall from the particle collisions. Due to the discontinuity in the gas MCD and thereby the delta function $\delta_S$ in the force density, one is able to define a {\it surface force},
\begin{equation}
    \Delta F^j
    = - \int_{\Delta \mathcal{V}} \partial_i T^{ij}_{\rm gas} {\rm d}\mathcal{V}
    = \left[ \left(T^{ij}_{\rm gas}\right)_{\rm in}- \left(T^{ij}_{\rm gas}\right)_{\rm out}\right] n^i \Delta S \ , \label{eq:Disc-Force}
\end{equation}
where the volume is chosen as an infinitesimally thin pillbox-shaped cylinder, oriented perpendicular to the wall, with its flat faces $\Delta S$ lying just inside and outside the boundary. Generally, this force can be decomposed into normal and tangential components. While the tangential part corresponds to a shear force, the pressure force is {\it defined} as the normal force acting on a unit boundary surface element, 
\begin{equation}
    p \equiv \frac{\Delta F^j n^j}{\Delta S} = \left[ \left(T^{ij}_{\rm gas}\right)_{\rm in}- \left(T^{ij}_{\rm gas}\right)_{\rm out}\right] n^i n^j = p_{K} + p_{I}\ . \label{eq:Pressure}
\end{equation}
The above equation shows that {\it the discontinuity of the MCD} exhibits a {\it direct mechanical effect}: the surface force acting on the corresponding plane. The superpotential contribution has zero divergence and is therefore continuous along the normal direction and drops out of the above difference. This reinforces the point that the MCD itself does not carry inherent mechanical significance---even though part of it has been called pressure. Only its discontinuous part, which contributes to the divergence, is capable of generating a surface force. This result is obviously consistent with the pressure-volume force for gases in thermodynamics, which contains both a kinetic part $p_{K}$ and an interaction part $p_{I}$, as in the Van der Waals equation. 

\begin{figure}[ht]
    \centering
    \includegraphics[width=0.5\linewidth]{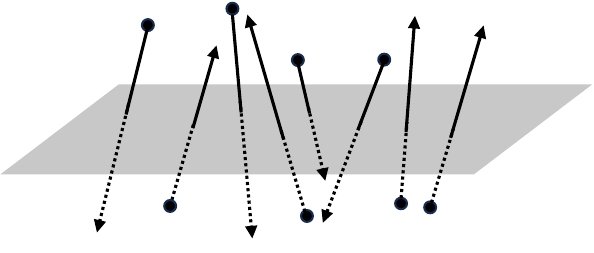}
    \caption{A fictitious plane inside a gas, through which the isotropic kinetic MCD represents a momentum flow transported by particles going from one side to the other. No physical force or mechanical effect is present in this process.}
    \label{fig:gas-flux}
\end{figure}

What about the forces inside the gas? In the case of ideal gases, the kinetic pressure does not correspond to any internal force. In fact, the physical meaning of the kinetic MCD or pressure $p_K$ within the gas is as follows: consider a hypothetical plane within the gas, as depicted in FIG.~\ref{fig:gas-flux}, across which particles continuously traverse from both sides. If we exclusively focus on one side, particles are effectively ``lost'' as they cross to the opposite side, but on average, they reappear with opposite momentum. Consequently, this results in an apparent net momentum flux into the region, which may be interpreted as an effective ``pressure'' exerted by the opposing side. However, in the absence of intermolecular interactions, there is no actual physical force involved. The concept of pressure here is completely virtual, manifesting only when a discontinuity is introduced, such as by inserting a physical surface or boundary. 

On the other hand, the interaction MCD arising from short-range forces can be interpreted as a genuine contact force within the gas. Consider again the same plane in FIG.~\ref{fig:gas-flux}, where the interaction MCD generated by one side of the gas $T^{\prime ij}_{I}$ decays rapidly near the plane. As the interaction range approaches zero, a discontinuity in $T^{\prime ij}_{I}$ is generated. The force exerted from one side onto the other can then be computed by integrating the divergence of the MCD across a thin volume enclosing the plane. In the zero-range force limit, the contribution to this force comes entirely from the MCD on the fictitious plane itself. Therefore, the interaction MCD $\delta^{ij}p_I$ indeed represents a normal surface force: it is positive (negative) for repulsive (attractive) interactions~\cite{van1873over}.

\subsection{Momentum Flow in Liquids: Compression Pressure and Viscous Forces}

Unlike gases where particles move freely, molecules in a liquid are restricted to move primarily around each other, allowing liquids to flow and conform to the shape of their containers while preventing them from expanding to fill the entire volume. Liquids have long been modeled as continuous media.  

The MCD of liquids contains both contributions from macroscopic kinetic motions (or flow) of elements and the internal forces,
\begin{align}
    T^{0i}_{\rm liquid}(\vec{r}) &= \rho v^i(\vec{r}) \ ,\\
    T^{ij}_{\rm liquid}(\vec{r}) &= \rho v^i(\vec{r}) v^j(\vec{r}) +\delta^{ij}p - \sigma^{ij}(\vec{r}) \ , \label{eq:MCD-Liquid}
\end{align}
where $\rho$ is the liquid density, $v^i(\vec{r})$ is the velocity field. As in a gas, there is an isotropic interaction pressure $p$ acting in the normal direction of any fictitious plane. In the static equilibrium without any external force, contributions from both thermal motion and interaction potentials tend to balance out, resulting in $p\sim 0$. A non-zero positive $p$ typically arises from the repulsive contact interactions when a liquid is subjected to an external surface or body compression force, as through gravity. The stress tensor $\sigma^{ij}$ is the viscous surface force due to the velocity gradient of the flow, 
\begin{align}
  \sigma_{ij} =\eta \left(\partial_i v_j +\partial_j v_i-\frac{2}{3} \delta_{ij} \partial_kv_k\right)+ \zeta\partial_kv_k  \delta_{ij}
 \ , 
\end{align}
where $\eta$ and $\zeta$ are the shear and bulk viscosity, respectively. In the absence of external forces, the conservation law $\partial_0 T^{0i}_{\rm liquid}+\partial_iT^{ij}_{\rm liquid}=0$ leads to the well-known Navier-Stokes equation inside the liquids. 

In the presence of external gravitational field in $z$ direction (chosen to be positive upward), one adds another contribution $T_g^{ij}$ to the total MCD, 
\begin{equation}
   T^{ij}_g = V_g (z) \delta^{ij} \ , 
\end{equation}
where $V_g(z) =\rho g(z-z_0)$ is the gravitational potential and $z_0$ can be chosen as the surface of the liquid. Now, the total MCD $T^{ij}_{\rm tot} = T^{ij}_{\rm liquid} + T^{ij}_g  = \delta^{ij}[p(z)+V_g(z)]$ is conserved in the liquid: $\partial_i T^{ij}_{\rm tot} = 0$. In the static equilibrium, the continuity equation reduces to 
\begin{equation}
   \frac{{\rm d}p(z)}{{\rm d}z} = - \frac{{\rm d}V_g(z)}{{\rm d}z}
\end{equation}
Therefore, $T^{ij}_{\rm tot}$ is again a superpotential and does not carry specific mechanical significance. Moreover, the external long-range gravity contributes a ``pressure potential'' which appears to cancel the actual pressure in the liquid. However, this ``cancellation'' occurs only at the force density level, is only complete modulo a constant term. Moreover, the pressure potential can be chosen entirely negative, like the boundary effect for gases.

In liquids, the frictional forces, arising from the shear and bulk viscosity, act to slow down the macroscopic flow while simultaneously generating heat. This is essentially an energy transfer between different scales: the macroscopic kinetic energy---represented by the drift velocity of liquid elements $v^i$---is converted into microscopic random thermal motions. Therefore, the concept of shear forces in liquids arises only in the consideration of a scale separation and dissipative mechanism.

\subsection{Momentum Flow in Solids: Stress Tensor for Contact Forces}

Unlike gases and liquids, solids are composed of atoms that are tightly bound due to strong inter-molecular interactions. Theory of elasticity~\cite{Landau:1986aog} was established to describe mechanical properties of solids, such as internal stresses and deformations, where they are again modeled as continuous media. 

The state of a continuous elastic material can be described by a displacement vector field $\vec{u}(\vec{r}) = \vec{r}'-\vec{r}$ at each point in space. Its MCD receives contributions from both the kinetic motion of elements as well as the interactions among them, 
\begin{align}
    T^{0i}_{\rm solid}(\vec{r}) &= \rho(\vec{r}) v^i(\vec{r}) \ ; \label{eq:MD-Solid}\\
    T^{ij}_{\rm solid}(\vec{r}) &= \rho(\vec{r}) v^i(\vec{r}) v^j(\vec{r}) - \sigma^{ij}(\vec{r}) \ , \label{eq:MCD-Solid}
\end{align}
where $\rho(\vec{r})$ and $v^i = \dot{u}^i$ are the local density and velocity at $\vec{r}$, $\sigma^{ij}(\vec{r})$ is a symmetric tensor representing the internal force effect. 

To explore the physics of momentum flow of internal forces $\sigma^{ij}(\vec{r})$, one may consider small oscillations and study the momentum continuity equation, which provides the dynamical equation, 
\begin{equation} 
    \rho\ddot{u}^{j}-\partial_{i}\sigma^{ij}=0\ , \label{eq:EoM-Solid}
\end{equation}
where $\ddot{u}^i$ is the acceleration and higher-order nonlinear terms have been omitted. Therefore, the divergence $\partial_i \sigma^{ij}=f^{j}$ is the internal force density acting on the solid elements. However, this equation does not uniquely determine the interaction MCD $\sigma^{ij}$ without further assumptions. For example, one can add a superpotential, such as $(\nabla^i \nabla^j - \delta^{ij} \nabla^2)F(\vec{r})$, without changing the above equation. 

Theory of elasticity offers a concrete model for the interaction MCD, the stress tensor, which is defined via Hooke's law as the contact forces within solids. In this framework, the interaction MCD flowing through a surface element, $\sigma^{ij}{\rm d}S^j$, directly corresponds to  the internal surface force acting on it. It is evident that $\sigma^{ij}$ must vanish outside the solids, thereby establishing a {\it boundary condition} to determine it from the force density.

Hooke's law states that internal stresses are related to the change in distance between adjacent points, quantitatively described by the strain tensor, $u_{ij} = \partial_{(i}u_{j)}$. It provides a linear relation for solids undergoing small deformations, 
\begin{equation} 
    \sigma_{ij}(\vec{r}) = K\delta_{ij}u_{ll}(\vec{r})+2\mu\left[u_{ij}(\vec{r})-\frac{1}{3}\delta_{ij}u_{ll}(\vec{r})\right] \ , \label{eq:Hooke}
\end{equation}
where $K$ and $\mu$ are respectively the bulk modulus and the shear modulus; both may depend on local parameters such as temperature, etc~\cite{Landau:1986aog}. In this model, the six components of $\sigma^{ij}$ are now expressed using three independent degrees of freedom associated with $u^i$. Consequently, the force density equation $\partial_i\sigma^{ij}=f^j$, together with appropriate boundary conditions, enables a complete determination of the stress tensor. 

In this linearized theory, an ideal static solid with no deformation satisfies $\sigma^{ij}=0$. Static deformations arise due to external forces, which may be applied either through boundary conditions or through long-range fields such as gravity and electromagnetism. In such cases, the MCD of solids in static equilibrium becomes $T^{ij}_{\rm solid}=-\sigma^{ij}$, which is no longer conserved, as external forces infuse momentum into the system---analogous to the role of container walls in gases and gravitational fields in liquids. Only upon inclusion of the external MCD, which in this context acts as a potential term, does the total MCD becomes conserved. However, this total MCD does not carry direct mechanical significance beyond representing the overall momentum flow. An explicit example of this will be discussed in Sec.~\ref{subsubsec:Charged_Sphere}. 

\subsection{Medium Dynamics as Classical Field Theory} \label{subsec:Phonon}

When treating solids as continuous media consisting of elements made up of a large number of molecules, we essentially adopt the classical particle dynamics where the momentum density and kinetic MCD are directly related to the mechanical motion of elements, and the interaction MCD arises from solving the continuity equation (Eqs. (\ref{eq:MD-Solid}) and (\ref{eq:MCD-Solid})). 

However, the particle-based description is of limited utility for formulating the momentum conservation law for the dynamics of continuous media, as the motion of individual particles is typically constrained. It is far more insightful to focus on the wave aspects, in which momentum is identified and transported along the direction of wave propagation. It is the local restoring contact interactions between elements that enable the propagation of disturbances. The wave momentum, which arising from the homogeneity of the medium, is essentially a {\it quasi momentum}. This distinction is clearly illustrated in the case of transverse wave, where particles are moving in a direction orthogonal to that of the wave propagation. Therefore, the particle and wave momenta are fundamentally different.

To describe the medium dynamics in terms of waves, one constructs a continuum field theory based on the displacement field $\vec{u}(\vec{r})$. The dynamics can then be derived from a suitably constructed Lagrangian density, 
\begin{equation}
    \mathcal{L}=\frac{1}{2}\rho\left(\partial_{t}\vec{u}\right)^{2}-\frac{1}{2}\lambda\left(\nabla\cdot\vec{u}\right)^{2}-\mu u_{ij}u_{ij} \ , \label{eq:L-Continuum}
\end{equation}
where $\lambda=K-2\mu/3$ and $\mu$ are Lam\'e coefficients, and the second and third terms are the elastic potential energy in the solid. The internal stress tensor can be defined through $\sigma^{ij}= -\partial {\cal L}/\partial u_{ij}$, recovering Hooke's law as in Eq.~(\ref{eq:Hooke}).

Using Noether's theorem for the translational symmetry of the medium, one can derive the canonical EMT for the field theory as,
\begin{align}
    T^{0i}&=-\rho\dot{u}_{k}\partial_{i}u_{k} \\
    T^{ij}&=\left(\lambda u_{ll}\delta_{ik}+2\mu u_{ik}\right)\partial_{j}u_{k}+\delta_{ij}\mathcal{L}=\sigma_{ik}\partial_{j}u_{k}+\delta_{ij}\mathcal{L}
\end{align}
Again, the above field momentum density and MCD are fundamentally different from the corresponding ones in the particle picture in Eqs. (\ref{eq:MD-Solid}) and (\ref{eq:MCD-Solid}). Here $T^{0i}$ arises from the propagation of disturbances through harmonic interactions in the medium, and the MCD represents the associated momentum flows through fields, just like the case of electromagnetism to be discussed later. Despite the different physical meanings, the momentum conservation still gives the same equation of motion apart from an extra factor, 
\begin{equation}
    \partial_{0}T^{0j} + \partial_{i}T^{ij} = - \left( \rho\ddot{u}_{k} - \partial_{i}\sigma_{ik}\right) \partial_{j}u_{k} = 0 \ .
\end{equation}
The divergence of this MCD no longer gives the physical force density acting on the solid elements, even though they are proportional to each other. Instead, this is a {\it quasi force density} that accounts for how the quasi momentum of the field is changed and conserved. In particular, in the above equation, the index $k$ represents the direction of the physical force density acting on the solid elements, whereas $j$ refers to the direction of the quasi force density on the field system. 

The two components of the stress tensor, the trace and traceless parts, indicate the two modes of elastic waves within solids: longitudinal and transverse waves, respectively. Therefore, a plane wave solution can in general be written as,
\begin{equation}
    \vec{u} = \vec{u}_{\perp}e^{-i\left(\omega_{\perp}t-\vec{k}\cdot\vec{r}\right)}+\vec{u}_{\parallel}e^{-i\left(\omega_{\parallel}t-\vec{k}\cdot\vec{r}\right)}
\end{equation}
where the speeds of sound of longitudinal and transverse waves $c_{\parallel},c_{\perp}$ are, 
\begin{equation}
    c_{\parallel}^{2}=\frac{\omega_{\parallel}}{k}=\frac{\lambda+2\mu}{\rho},~~~ c_{\perp}^{2}=\frac{\omega_{\perp}}{k}=\frac{\mu}{\rho} \ .
\end{equation}
The time-averaged MCD for the above place wave solution is, 
\begin{equation}
    \left\langle T^{ij} \right\rangle= \frac{1}{2} \rho(c_{\parallel}^2\vec{u}_{\parallel}^{2} + c_{\perp}^2\vec{u}_{\perp}^{2} )k^{i}k^{j}
\end{equation}
Again the MCD represents a quasi momentum flow in the direction of wave propagation, rather than the kinetic momentum flow of and forces on the medium particles.

The lesson we learn from the above example is that trying to recover the mechanical system---if it exists---behind a field theory is nontrivial, and is even unnecessary. Instead, in a classical field theory, one focuses on the forces and momenta associated with waves or the propagating of field disturbances, rather than observables of the underlying media supporting them. In fact, the physical forces of the medium, encoded in the quadratic terms of Eq.~(\ref{eq:L-Continuum}), are not interpreted as the contact forces in the field theory. Instead, they generate (quasi) momentum flow through spatial gradients of the fields, analogous to the momentum transport by particles. Specifically, while the contact force in a physical medium is so crucial, the elastic waves in the linearized theory are non-interacting and free-propagating with three eigenmodes, and there is no field force acting on them!

Therefore, our attitude towards the systems of classical fields, as far as momentum density and forces are concerned, is that the mechanics of any possible underlying media is not within reach. This is the case, for example, in the emergent classical field theories in statistical mechanics, such as the order parameters of phase transitions in many-body systems, for which (quasi) momentum conservation law may be studied. There are also classical field theories of electromagnetism and gravity for which Aether does not exist. The momentum is transported in these systems either through waves, or the non-dynamical degrees of freedom, i.e., gradients of static fields. The short-range contact forces and mechanical pressure cease to be natural concepts, except in nonlinear field theories where they re-emerge through the classical limit of quantum many-body physics, as we will discuss in the next section. 

\subsection{Momentum Flow in Electromagnetism: Long-Range Force through Fields} \label{subsec:E&M}

Everyday experiences tell us that forces are through contact interactions, as manifested in the systems in the previous subsections. However, classical gravity and electromagnetism are examples of long-range forces acting between objects separated by a distance. For these systems, the local conservation of momentum is intriguing, as it intends to picture {\it how momentum continuously flows through empty space} among particles. Classical electric and magnetic fields have been proposed to facilitate interactions, and the flow of momentum as well. At the level of electrostatics and magnetostatics, the field theory language simply represents a choice or a model to explain forces. Ultimately, the legitimacy of the interpretation lies in physical effects of the associated dynamics as well as coupling to gravity. 

\subsubsection{MCD of Coulomb Forces Through Electric Fields}

Discussions in Sec.~\ref{subsec:NRMCD} can be applied straightforwardly to a system of charges interacting under Coulomb's law. Solutions for the interaction MCD come from Eq.~(\ref{eq:MCD-Int}) with $\vec{F}_a$ replaced by Coulomb forces. Although there is an infinite number of possible choices for $T^{ij}_I$, one solution stands out for reasons beyond the mechanics of the system itself: instead of action-at-a-distance, the electric force was postulated to be conducted through an intermediary, the electric fields $\vec{E}(\vec{r})$ present everywhere in space. This allows for restoring locality of interactions, $\vec{F}(\vec{r})=q \vec{E}(\vec{r})$. 

In terms of local momentum conservation, the fields serve as a vehicle to transport momentum among charged objects, generating the local forces. The interaction MCD $T^{ij}_{I}(\vec{r}) \equiv T^{ij}_{E} (\vec{r})$ is, 
\begin{align}
    T^{ij}_{E} (\vec{r})
    &= \epsilon_{0}\left[\frac{1}{2}\delta^{ij}\vec{E}^{2}(\vec{r})-E^{i}(\vec{r})E^{j}(\vec{r})\right] \\ 
    &= \epsilon_0\left[\frac{1}{2}\delta^{ij}\nabla^k A^0 \nabla^k A^0 - \nabla^i A^0\nabla^j A^0\right]\ , \label{eq:MCD-E}
\end{align}
where $\epsilon_0$ is the vacuum permittivity, and $A^0(\vec{r})$ is the scalar field defined through $E^i\equiv-\nabla^i A^0$. Supposing that there is a static charge distribution $\rho(\vec{r})$ generating an electric field $\vec{E}(\vec{r})$, the divergence of the interaction MCD is given by, 
\begin{equation}
    \partial_i T^{ij}_{ E}(\vec{r}) = - \rho(\vec{r}) E^j(\vec{r}) = - {\cal F}^j(\vec{r})\ ,
\end{equation}
where ${\cal F}^j$ is the Coulomb force density exerted on the charge at $\vec{r}$ from other parts of the distribution. Notably, though the electric fields are present everywhere in space, forces only appear where there are charges, and there is no force among the (free) fields themselves. In fact, one can eliminate $\vec{E}$ through Gauss's law, $\nabla\cdot \vec{E} = \rho(\vec{r})/\epsilon_0$, and express $T^{ij}_I$ directly in terms integrals of the charge distribution without the electric field at all. There is no loss of physics so long as gravity is not involved. 

A peculiarity of the static electric field in field theory can be seen through a dynamical free scalar field $\phi(\vec{r},t)$ with the Lagrangian density, 
\begin{equation}
    {\cal L} = \frac{1}{2}(\partial_t \phi)^2 - \frac{1}{2} (\nabla \phi)^2 \ .
\end{equation}
This Lagrangian density will generate an MCD written as,
\begin{equation}
    T^{ij} = \partial^i \phi \partial^j \phi + \delta^{ij}{\cal L} \ .
\end{equation}
In the static case, it differs by a minus sign from Eq.~(\ref{eq:MCD-E}). This is due to the fact that $A^0$ is the fourth component of a Lorentz vector. 

In the following sub-subsections, through examples of two charges, charged plate in an external field, and a continuous medium, we demonstrate some special aspects of $T^{ij}_E$ in relation to forces and pressures. 

\subsubsection{Momentum Flows in Two-Charged-Particle System}

\begin{figure}[ht]
    \centering
    \includegraphics[width=1\linewidth]{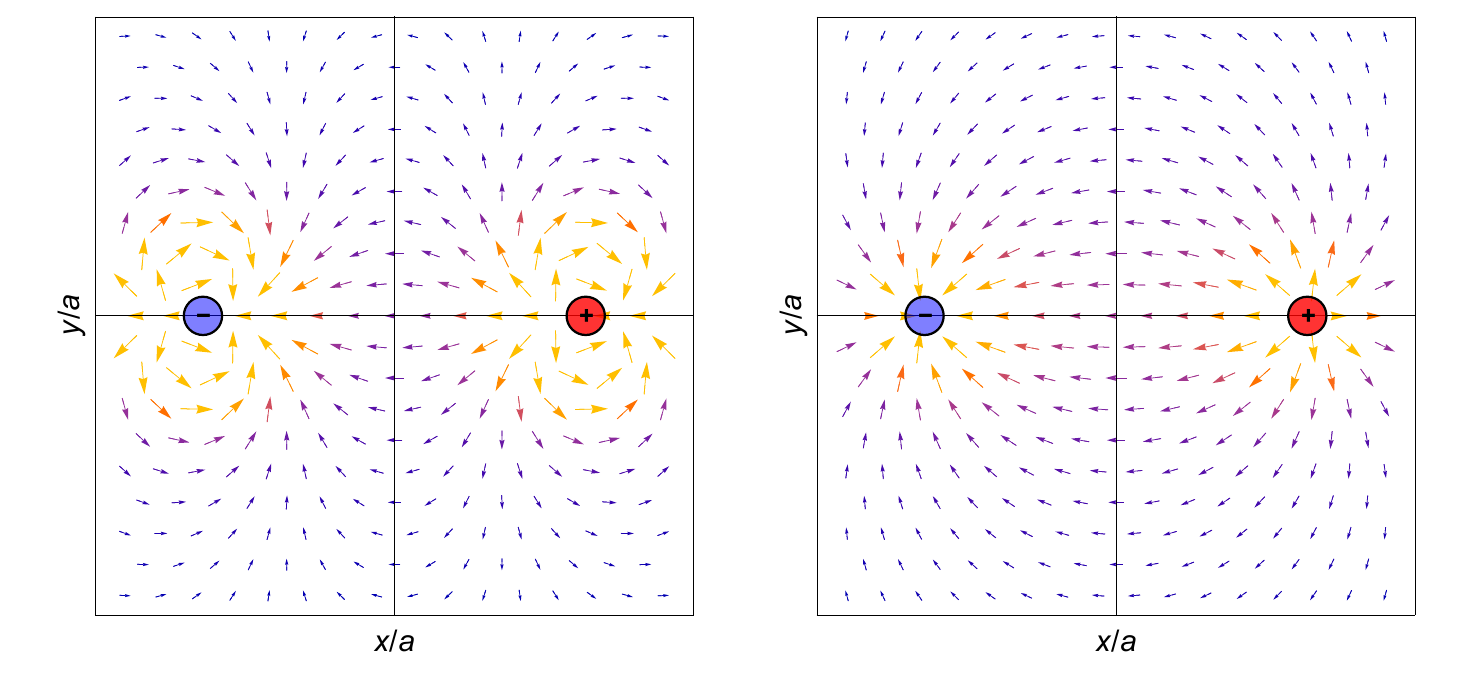}
    \caption{The flow of momentum $p_x$ in the $x-y$ plane for two opposite electric charges located at $(\pm a,0,0)$, indicating the Coulomb force between them. Total (left) and interference (right) MCDs have the same mechanical effects although differed by the self-field MCDs of charges equivalent to a superpotential.}
    \label{fig:EM-dipole}
\end{figure}

Even in the electrostatic theory, the momentum flow accounting for the electric force does not have to be unique. For instance, consider an electric dipole with two opposite charges $\pm q$ located at $\pm\vec{r}_0$, each generating one electric field $\vec{E}_{+}$ and $\vec{E}_{-}$. Since the Maxwell MCD is quadratic in the total electric field $\vec{E} = \vec{E}_{+} + \vec{E}_{-}$, it can be split into three terms,
\begin{equation}
     T_{E}^{ij} = T^{ij}_{+} + T^{ij}_{-} + T^{ij}_{\rm int} \equiv T_{\rm self}^{ij} + T^{ij}_{\rm int}\ ,\label{eq:twocharge}
\end{equation}
where $T^{ij}_{\pm}$ are the MCDs for self electric field of two charges and are collectively denoted as $T_{\rm self}^{ij}$, and $T^{ij}_{\rm int}$ is the interference contribution given as, 
\begin{equation}
    T^{ij}_{\rm int} = \epsilon_0 ( \delta^{ij}\vec{E}_{+}\cdot\vec{E}_{-}-E_{+}^i E_{-}^j-E_{+}^jE_{-}^i ) \ .
\end{equation}
While $T_{\rm self}^{ij}$ is conserved itself, the divergence of $T^{ij}_{\rm int}$ gives the electric force density, 
\begin{align}
    \partial_i T^{ij}_{\rm self} = 0\ ,~~~\partial_i T^{ij}_{\rm int} = - ( +q E_{-}^j ) \delta^{(3)} (\vec{r}-\vec{r}_0) - ( -q E_{+}^j) \delta^{(3)}(\vec{r}+\vec{r}_0) \ .
\end{align}
The above internal force density will be balanced by external forces to keep the two charges static. Therefore, $T^{ij}_{\rm self}$ can be regarded as a superpotential in Eq.~(\ref{eq:twocharge}): though the total and interference electric MCDs flow the momentum in different ways, as shown in FIG.~\ref{fig:EM-dipole}, they generate the same mechanical effects, i.e., Coulomb forces between charges. More generally, one is free to add arbitrary divergence-free terms without changing the interaction effects. However, for coupling to gravity,  the self-field MCDs of charged-neutral systems is critically important to cancel the long-distance effects from interference MCD and ensure a well-behaved convergence, as in the hydrogen atom to be discussed below. 

\subsubsection{Momentum Flow in a Charged Solid Sphere} \label{subsubsec:Charged_Sphere}

Microscopically, stress tensors in liquids and solids arise from the MCD of Coulomb interactions among electrons and positively-charged nuclei. Using quantum mechanics, one can calculate the effective interactions between atoms/molecules. When forming solids, individual atoms/molecules vibrate around their equilibrium positions of the interaction potentials like a network of 3D oscillators. The effective MCD or stress tensor, $\sigma^{ij}$, is solved from the force density distributions in the long-wavelength limit (or continuous medium) of vibrations. It is quite clear that $\sigma^{ij}$ is not an average of the original MCD $T^{ij}_{K} +T^{ij}_{E}$ at the fundamental level, although their corresponding force densities are definitely related by quantum mechanical as well as statistical averages. Their difference might be a superpotential term. Therefore, it is almost certain that the effective $T^{ij}$ of the solids and liquids will not be the one that couples to gravity, due to the absence of the explicit notion of electric fields. 

To illustrate the above point, let us consider a model of a point-like charge: a uniformly-charged sphere of radius $R$ and charge density $\rho$. One might assume that the sphere is a continuous medium made of some type of elastic material. Its mechanical stress tensor balances the static electric force from the charge density embedded in it. By explicitly solving the balance equation with Hooke's law~\cite{Landau:1986aog}, 
\begin{equation}
    \partial_i \sigma ^{ij}  = - \rho E^j  \ , \label{eq:stress}
\end{equation}
one obtains the spherically symmetric displacement vector (deformation) as,
\begin{equation}
    \vec{u}(\vec{r})=\frac{\rho^{2}}{2\epsilon_{0}}\frac{1}{3K+4\mu}\left[\frac{1}{3}\left(1+\frac{8}{15}\frac{\mu}{K}\right)R^{2}-\frac{1}{5}r^{2}\right]\vec{r} \ ,
\end{equation}
where $K$ and $\mu$ are the medium parameters introduced before. The above deformation generates the stress tensor through Hooke's law,
\begin{equation}
    -\sigma^{ij} = \left\{ -\rho\left[V(r)-V(r)\right]\delta^{ij}+\frac{\mu}{3K+4\mu}\frac{\rho^{2}}{20\epsilon_{0}}\left(\nabla^{i}\nabla^{j}-\delta^{ij}\nabla^{2}\right)(r^{2}-R^{2})^{2}\right\} \theta_{\cal V} \ ,
\end{equation}
which obeys the boundary condition that $\sigma^{ij}$ vanishes outside the sphere. Here we introduced the electric potential $V(r)$ and collected the rest as a divergence free term. Notably, $\sigma^{ij}$ is the mechanical pressure and shear force inside the sphere as the response to the long-range electric force. When $\mu=0$, $\sigma^{ij}=\rho [V(r)-V(R)]\theta_{\cal V}\delta^{ij}$ is a simple pressure-like term, giving rise to a negative pressure force $p(r)=-\rho [V(r)-V(R)]\theta_{\cal V}$ that pushes in. This is an actual surface force even though $V(r)$ is a potential. If the electric force is replaced by gravitational interactions, one has a positive pressure force, pushing the medium out. 

On the other hand, the MCD from the electric field, satisfying exactly the same equation (apart from a minus sign) as Eq.~(\ref{eq:stress}), is given as,
\begin{align}
    T^{ij}_{E}(r) =& \left\{ \rho\left[V(r)-V(r)\right]\delta^{ij}-\frac{\rho^{2}}{72\epsilon_{0}}\left(\nabla^{i}\nabla^{j}-\delta^{ij}\nabla^{2}\right)r^{4}-\frac{\rho^{2}}{2\epsilon_{0}}R^{2}\delta^{ij}\right\} \theta_{\cal V} \nonumber \\
    &+ \left[ \dfrac{\rho^{2}}{72\epsilon_{0}}(\delta^{ij}\nabla^{2}-\nabla^{i}\nabla^{j})\dfrac{R^{6}}{r^{2}} \right] \left(1-\theta_{\cal V}\right) \ ,
\end{align}
The sphere is in mechanical balance and therefore the total MCD is conserved,
\begin{equation}
    T_{\rm tot}^{ij} = T^{ij}_{ E} - \sigma^{ij}; ~~~~~~ \partial_i T_{\rm tot}^{ij}=0 \ ,
\end{equation}
which can be easily checked from the above result. On the other hand, $T^{ij}_{\rm tot}$ is just the difference of two MCDs that satisfy the same Eq.~(\ref{eq:stress}) but with two different boundary conditions, and therefore amounts to a superpotential. While $\sigma^{ij}$ has the interpretation as the mechanical stress, $T^{ij}_{E}$ does not. The total MCD satisfying Eq.~(\ref{eq:von-Laue}) has no information about pressure and force: there is no net force density at any point in the sphere. At a more fundamental level, $T_{\rm tot}^{ij}$, entirely of the electric origin, is not the one that couples to gravitational fields. 

One may think of the point charge, such as the electron, as the limit of a uniformly charged sphere with a radius $R\to 0$ and charge density $\rho\to \infty$. The MCD for the point-like charge is singular at the origin. Regularizing the UV divergence while maintaining the translational symmetry requires introducing a mechanism of stability for the point-like charge~\cite{poincare_sur_1906}, similar to $\sigma^{ij}$ above, which is tacitly assumed in QFTs. This external force can also be explained as a UV subtraction contribution which makes the positive definite $T^{ii}_{E}(r)$ integrates to zero after regularization as in Eq.~(\ref{eq:von-Laue}). 

\subsubsection{Electromagnetic MCD Through Propagation of Waves}

Although static electromagnetic fields appear not indispensable, Maxwell's theory introduces dynamical components and predicts the electromagnetic radiations without sources. Therefore the MCD for E\&M fields in general receives two separate contributions: the static interactions and E\&M waves. 

Spacetime translational symmetry leads to a conserved EMT of the E\&M fields. Through the Belinfante-Rosenfeld improvement, one obtains a gauge-invariant and symmetric EMT, which is believed to be a source for the warpage of the spacetime in General Relativity. The momentum density and MCD are, 
\begin{align}
    T^{0i}_{\rm em} (\vec{r}) &= \frac{1}{\mu_0 c} \left[ \vec{E}(\vec{r})\times\vec{B}(\vec{r}) \right]^i \ ,  \\
    T^{ij}_{\rm em} (\vec{r}) &= \epsilon_{0}\left[\frac{1}{2}\delta^{ij}\vec{E}^{2}(\vec{r})-E^{i}(\vec{r})E^{j}(\vec{r})\right]+\frac{1}{\mu_{0}}\left[\frac{1}{2}\delta^{ij}\vec{B}^{2}(\vec{r})-B^{i}(\vec{r})B^{j}(\vec{r})\right] \ , 
    \label{eq:MCD-EM}
\end{align}
where $E^i, B^i$ are the total electric and magnetic fields, respectively. One may choose the Coulomb gauge, $\nabla \cdot \vec{A}=0$, such that, 
\begin{align}
    \vec{E} = -\nabla \phi -\dot{\vec{A}}_\perp; ~~~ \vec{B} = \nabla\times \vec{A}_\perp \ . 
\end{align}
The non-dynamical, longitudinal part of $\vec{E}$ does contribute both to the momentum density as well as the MCD. Moreover, in a static $\vec{B}$ field, there is also a non-vanishing MCD. 

The E\&M radiation contributions can be described using the language of waves, where the momentum flow can be viewed as a direct transport through wave propagation. Consider a plane wave in the direction of $\vec{k}$, 
\begin{equation}
    \vec{A}_\perp = \vec{A}_{\perp 0} e^{-i(\omega t- \vec{k}\cdot \vec{r})} \ .
\end{equation}
The corresponding MCD is, 
\begin{equation}
    T^{ij}_{\rm em} = \frac{1}{2\mu_0}|\vec{A}_{\perp 0}|^2 k^i k^j \ , 
\end{equation}
where we have averaged over the rapid oscillations in time. The above MCD has exactly the same form as Eq.~(\ref{eq:MCD-idealgas}) of a free gas with momentum in the direction of $\vec{k}$,  flowing with a velocity $v^i \sim k^i$ and energy density $\sim |A_{\perp 0}|^2/2\mu_0$. The absorption or scattering of E\&M radiations can produce a radiation pressure~\cite{lebedev1883experimental,nichols1903pressure}.


Meanwhile, due to the relativistic nature of E\&M fields, there is no frame-independent, gauge-invariant separation between the static and dynamical parts of the E\&M MCD when both are present. In a bound relativistic system, there is a mixture of both. Again, the static contributions to MCD, either electric or magnetic, have no physical or mechanical stress interpretation other than their divergences generate the static force densities. 

\subsection{Non-Linear Field Theories and Soliton Solutions}

We have considered only the simplest linearized classical field theories so far. In the static case, the fields usually require external sources or boundary conditions to have non-zero values. However, if there are terms higher than quadratic in the Lagrangian, the fields will interact and these non-linear interactions can support non-zero values of fields in the lowest energy state. One of the most famous examples is the Landau-Ginzburg theory for phase transitions, for which the MCD for symmetry-breaking states is, 
\begin{equation}
    T^{ij}_0 = -\delta^{ij}{\cal H}_I(\phi_0)
\end{equation}
where ${\cal H}_I$ is the interacting Hamiltonian density of the order parameters $\phi_0$, which is negative in ordered state, resulting in a positive pressure-like term. However, $T^{ij}_0$ has no physical effects unless a discontinuity is created.  One important example is the Higgs theory in electroweak symmetry breaking. 

In the case of non-trivial topologies, finite energy solutions resembling composite particles can appear. The simplest example is a $\phi^4$ theory, which permits stable soliton solutions in 1+1 dimension. In this case, we may talk about forces between different parts of ``virtual'' waves. There have been models proposed to describe nucleons as solitons of static pion fields. For example, in the Skyrme model, the nucleon is identified as the topological solution of the non-linear sigma model of pions~\cite{Skyrme:1962vh,Adkins:1983ya,Cebulla:2007ei,GarciaMartin-Caro:2023klo,GarciaMartin-Caro:2023toa}. The Lagrangian density for this Skyrme model of nucleon is, 
\begin{equation}
    \mathcal{L}=\frac{1}{16}f_{\pi}^{2}{\rm Tr}(\partial_{\mu}U\partial^{\mu}U^{\dagger})+\frac{1}{32e^{2}}{\rm Tr}([U^{\dagger}\partial_{\mu}U,U^{\dagger}\partial_{\nu}U][U^{\dagger}\partial^{\mu}U,U^{\dagger}\partial^{\nu}U])
\end{equation}
where $f_\pi$ is a pion decay constant with mass dimension 1, and $e$ is a dimensionless coupling. For static field solutions, one adopts the Skyrme (or “hedgehog”) ansatz which couples the space and isospin, 
\begin{equation}
    U=\exp\left[iF(r)\vec{\tau}\cdot\hat{r}\right] = \cos[F(r)]+i(\vec{\tau}\cdot\hat{r})\sin[F(r)]\equiv\phi_{0}+i\tau_{a}\phi_{a}
\end{equation}
where $a = 1,2,3$ is an isospin index, and $\vec{\tau}$ is a 2-D isospin matrix. 
Notably, $\phi_0$ is not an independent variable since $\phi_0^2 + \phi_a\phi_a=1$. In fact, $\phi_{a=1,2,3}$ corresponds to three pions. 

\begin{figure}[ht]
    \centering
    \includegraphics[width=0.65\linewidth]{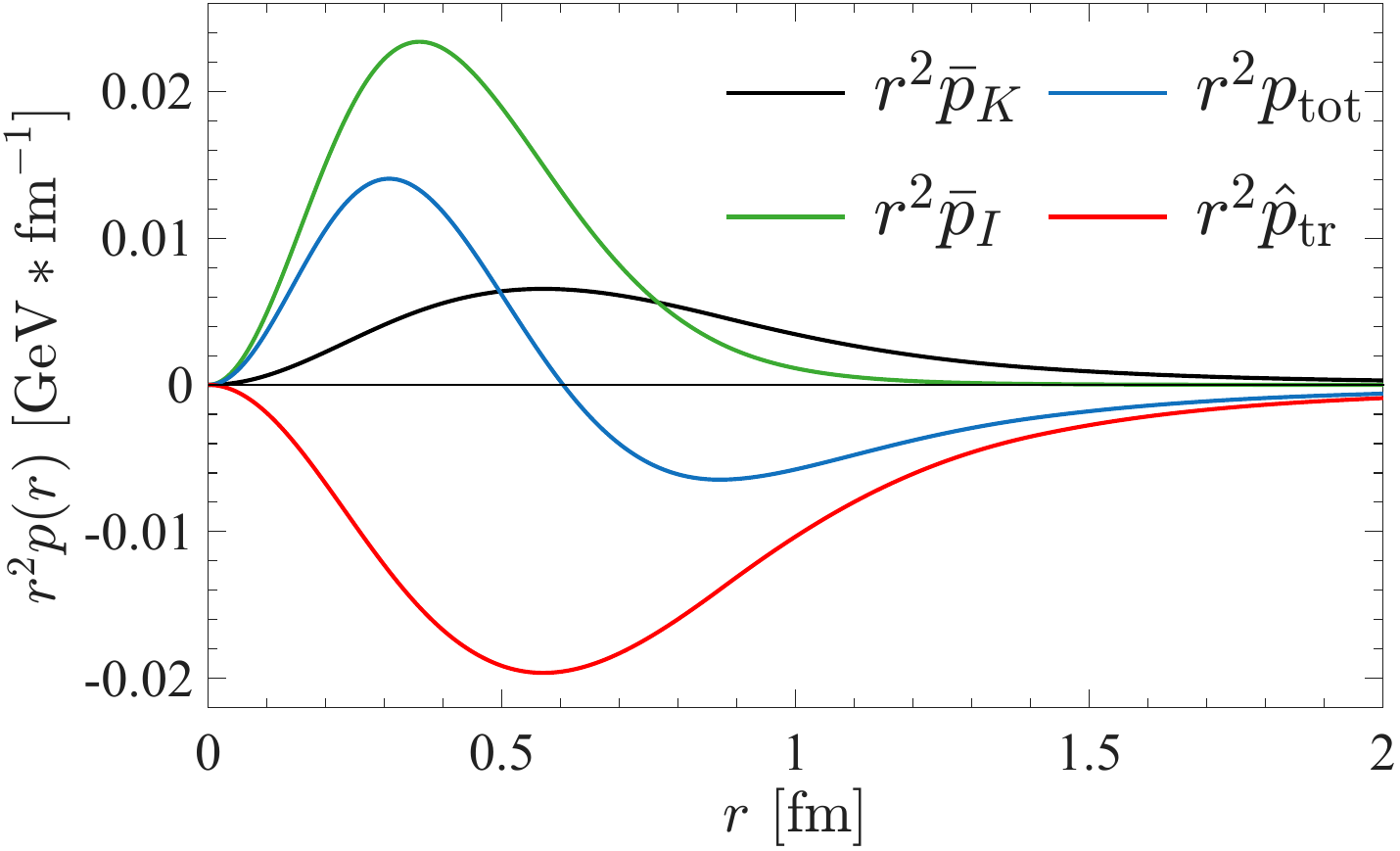}
    \caption{Traces of 4D traceless kinetic (black), interaction (green), trace (red) and the total (blue) MCDs $T^{ii}$ (top)~\cite{Cebulla:2007ei} within pion fields of the Skyrme model. 
    }
    \label{fig:MCD-Skyrme}
\end{figure}

To explore the physics of MCD in the Skyrme model, one may assume small perturbations for pion fields $\phi_a \ll 1$. Using this ansatz, the Lagrangian density is an interaction theory of pion fields, 
\begin{equation}
    \mathcal{L}=-\frac{1}{8}f_{\pi}^{2}(\partial_{i}\phi_{a}\partial_{i}\phi_{a})-\frac{1}{4e^{2}}[(\partial_{i}\phi_{a}\partial_{i}\phi_{a})^{2}-(\partial_{i}\phi_{a}\partial_{j}\phi_{a})(\partial_{i}\phi_{b}\partial_{j}\phi_{b})]
\end{equation}
where repeated indices $i,j,a$ are summed over. This contains both the quadratic terms, just like in E\&M, but also has quartic term which gives rise to non-linear but local interactions. Using Noether's theorem, one finds the field EMT is given as,
\begin{equation}
    T^{\mu\nu} = \frac{f_{\pi}^{2}}{4} (\partial^{\mu}\phi_{a}\partial^{\nu}\phi_{a}) + \frac{1}{e^{2}}(\partial_{\rho}\phi_{b}\partial^{\rho}\phi_{a}\partial^{\mu}\phi_{b}\partial^{\nu}\phi_{a}-\partial_{\rho}\phi_{b}\partial^{\rho}\phi_{b}\partial^{\mu}\phi_{a}\partial^{\nu}\phi_{a})-g^{\mu\nu}\mathcal{L}
\end{equation}
Therefore, one can decompose the above EMT into the 4D trace and traceless parts, and identify separate contributions to the MCD, including the kinetic and interaction parts, 
\begin{align}
    \bar{T}^{ij}_{K} &= \frac{f_{\pi}^{2}}{4}\left[(\partial_{i}\phi_{a}\partial_{j}\phi_{a})-\frac{1}{4}\delta^{ij}\partial_{k}\phi_{a}\partial_{k}\phi_{a}\right] \\
    \bar{T}^{ij}_{I} &= \frac{1}{e^{2}}(\partial_{k}\phi_{b}\partial_{k}\phi_{b}\partial_{i}\phi_{a}\partial_{j}\phi_{a}-\partial_{k}\phi_{b}\partial_{k}\phi_{a}\partial_{i}\phi_{b}\partial_{j}\phi_{a})+\delta_{ij}\mathcal{L}_{V} \\
    \hat{T}^{ij} &= \frac{f_{\pi}^{2}}{4}\left[-\frac{1}{4}\delta^{ij}\partial_{k}\phi_{a}\partial_{k}\phi_{a}\right]
\end{align}
where $T^{ij}_{K}\equiv \bar{T}^{ij}_{K} + \hat{T}^{ij}$ is the free virtual wave contributions. Notably, the 4D trace part comes entirely from the kinetic part of the scalar field and provides the scale of the system. The interaction MCD, which is traceless, contains four pion-fields derivative coupling. One can verify that the momentum continuity equation reproduces the Euler-Lagrange equations of motion for static fields.

Using the solution of Skyrme profile $F(r)$ with Skyrme parameters $f_\pi=129~{\rm MeV}$ and $e=5.45$ from~\cite{Adkins:1983ya} for the above Skyrme ansatz, we plot the 3D traces of the MCD $T^{ii}$ in FIG.~\ref{fig:MCD-Skyrme}. While the 4D traceless parts $\bar{T}^{ii}_K$ (not a pressure according to our previous discussion) and $\bar{T}^{ii}_I$ are positive and contribute mainly near the center, the 4D trace part $\hat{p}_{\rm tr}=\hat{T}^{ii}/3$ is large and negative, and dominates the large $r$-region, resulting in a negative total MCD $T^{ii}_{\rm tot}$. This is consistent with recent results from the modified Skyrme model~\cite{Fujii:2025aip,Tanaka:2025pny}, here without introducing another meson field. This 4D-trace contribution does not have a corresponding momentum density and therefore the conduction of the momentum flow does not include particle transport. It may be viewed as a pressure contribution in the Skyrme model. The total kinetic MCD $T^{ii}_{K}$ would be negative definite, i.e., the virtual-wave momentum flows in the opposite direction to the wave vector! This part of the MCD is similar to that of static electric field, but with an important sign difference.

\section{MCDs and Pressure/Forces in Quantum Systems}\label{sec:MCD-Quantum}

In this section, we focus on the MCDs and their physical effects in various quantum systems. Unlike statistical systems in which thermal averages naturally generate the concept of pressure, quantum averages are facilitated with probability amplitudes and generally exhibit ordered kinetic momentum flow due to wave-mechanical properties, thereby the concept of pressure-volume work does not naturally arise. Instead, the energies of such systems show more complicated dependency on the geometry of the volumes. 

We also investigate interacting quantum systems, such as the hydrogen atom (H-atom) and the Thomas-Fermi model for large atoms. Apart from the kinetic MCD associated with particle motions, the interaction MCDs are from long-range Coulomb forces. Here our conclusions about MCDs in classical electromagnetism still apply. Namely, the interaction MCDs are in general not interpretable as pressure and not directly surface forces, and only their divergences give rise to the force density. Moreover, not all terms in the interaction MCDs have force effects---the MCDs associated with the self electric fields do not have mechanical effects. 

Systems with a large number of bosons occupying the same single-particle quantum wave function are particularly interesting. This Bose-Einstein condensate can be described by a classical (complex scalar) field theory. When non-linear local field interactions are present, one finds that it describes short-range interactions between bosons. We then can recover the interaction MCD as a mechanical stress tensor in this system. 

Isotropy is restored in media containing large number of identical fermions, governed by quantum statistics. High-density fermionic gases homogenize and isotropize the ordered flow by occupying numerous higher-energy single-particle states, thereby recovering the concept of pressure-volume work. This is the case for nuclear matter or a large nucleus where the short-range interactions also mean the interaction pressure is a contact surface force. 

\subsection{Free Particles in a Box: Kinetic MCDs and Local Pressure}

In quantum mechanics, the average of the kinetic motion of particles is different from the statistical average in classical thermal systems. Moreover, the concept of volume, which is crucial to pressure-volume work, is often ambiguous because a clear boundary does not exist for bound states (this also happens of course for atmosphere around the earth). For instance, bound-state wave functions decaying smoothly to zero at spatial infinity, while still preserving a finite normalization. Nonetheless, we can still define local pressure through the pressure operator. In particular, the M.I.T. bag model for the nucleon uses the local pressure balance at the boundary. 

\subsubsection{Quantum Pressure vs. Ordered Motion}

Quantum pressure is an interesting concept that is best shown through a particle moving in a one-dimensional infinite square-well potential, 
\begin{equation}
    V\left(x\right)=
        \begin{cases}
            0 & 0<x<L \\
            +\infty & {\rm otherwise}
        \end{cases} \ .
\end{equation}
The quantum eigenstates and eigenvalues are in textbooks,
\begin{align}
    \varphi_{n}\left(x\right)  =\sqrt{\frac{2}{L}} \sin\frac{n\pi x}{L} \ ; ~~~ 
E_{n}  =\frac{n^{2}\pi^{2}\hbar^{2}}{2mL^{2}} \ ,
\end{align}
where $m$ is the mass of the particle and $n=1,2,3,...$ labels the energy levels that are dependent on the length/volume of the box. In this case, $L$ is the volume of the system. This dependence shows that the particle exerts a force on the boundary of the well, 
\begin{equation}
    F_{n}=-\frac{{\rm d}E_{n}}{{\rm d}L}= + \frac{n^{2}\pi^{2}\hbar^{2}}{mL^{3}} \ . 
\end{equation} 
The positive sign means that the force is against the well. The wave functions have derivative discontinuities and tend to lower their energies by diffusing over space, which is quite similar to a classical ideal gas in the box. 

Likewise, in the well-known Casimir effect, two infinite conductors separated by a distance $L$ forbid the presence of long-wavelength photons, leading to a volume-dependent energy gap $\Delta E(L)$ between the inner and outer vacua. This dependence implies the attractive force/pressure on the two plates~\cite{Casimir:1948dh,Lamoreaux:2005gf}, 
\begin{equation}
   p = \frac{F_{\rm Cas.}}{A} = - \frac{\partial \Delta E(L)}{\partial L} = -\frac{\pi^{2}\hbar c}{240L^{4}} \ ,
\end{equation}
where $A$ is the surface area of the plate. 

However, one sees an immediate difficulty when generalizing the above example to a 2D rectangular box with lengths $L_x$ and $L_y$, in which the energy eigenvalues become,
\begin{equation}
    E_{n_x,n_y}  =\frac{\pi^{2}\hbar^{2}}{2m}\left(\frac{n^2_x}{L_x^2}+ \frac{n^2_y}{L_y^2}\right) \ ,
\end{equation}
where $n_x, n_y=1,2,...$. The dependence on the geometry of the system is now $L_x$ and $L_y$ separately, rather than the volume, ${\cal V}=L_xL_y$. This is because the quantum states are waves that exhibit an ordered motion in two separate directions, different from the random thermal motion in gases. A quantum particle obviously does not fill the entire container homogeneously, or in other words, the local isotropy preserved by classical ideal gases, are broken. Therefore, the concept of familiar pressure-volume work, $p= -\partial E/\partial {\cal V}$, becomes invalid. 

\subsubsection{Quantum Kinetic MCD and Vacuum Pressure in M.I.T. Bag Model}

The pressures and forces in quantum systems can be studied through the quantum version of momentum conservation and MCDs. The most natural definition starts from the kinetic MCDs for classical particles in Eq.~(\ref{eq:MCD-Particle}), and promotes the momentum $\vec{k}_a$ and position $\vec{r}_a$ into operators by adding hats, 
\begin{equation}
    \hat{T}^{ij}_{K}(\vec{r}) = \sum_a \frac{1}{m_a} \hat{k}^i_a  \hat{k}^j_a\delta^{(3)}(\vec{r}-\hat{\vec{r}}_a)\bigg{|}_{\rm sym} \ , \label{eq:MCD-QuantumKinetic}
\end{equation}
where we need to completely symmetrize all operators to keep the product hermitian. For an arbitrary quantum state $\left|\varphi\right\rangle$, the expectation value is~\cite{Ji:2022exr, Freese:2024rkr}, 
\begin{equation}
    T^{ij}_{\varphi-{K}}(\vec{r})
    \equiv \langle \varphi|\hat{T}^{ij}_{K}(\vec{r})|\varphi\rangle =\frac{\hbar^{2}}{4m}\left[\nabla^i\varphi^{*}(\vec{r})\nabla^j\varphi(\vec{r})-\varphi^{*}(\vec{r})\nabla^i\nabla^j\varphi(\vec{r})+{\rm c.c.}\right] \ ,
\end{equation}
where c.c. stands for complex conjugate. In the presence of gauge fields, the momentum operator $\hat{k}^{i}$ and the partial derivative should be replaced by the canonical ones. 

Thus the quantum mechanics of a single particle is analogous to the classical wave mechanics of a complex field. In fact, if one writes down the corresponding Lagrangian density in terms of the quadratic dependence of wave functions, the global U(1) symmetry leads to the probability conservation, and the translational symmetry leads to the same kinetic MCD shown above. When there is a position-dependent potential, the symmetry breaking leads to additional terms in the MCD, which can be obtained from classical $T^{ij}_I$ through quantization (see next subsection).

By applying the above operator to the 1D infinite well, one obtains the kinetic MCD in the eigenstates, 
\begin{equation}
    T^{xx}_{n-{K}}\left(x\right)  =\frac{n^{2}\pi^{2}\hbar^{2}}{mL^{3}} \theta_{\cal V} \ .
\end{equation}
The momentum flow is a constant inside the box where there is no force, and has a discontinuity at the boundary because the wave function vanishes outside. Therefore, the kinetic MCD is not conserved at the boundary, indicating a confining force by the wall on the particle. The MCD near the wall is exactly the force acting on it, $T^{xx}_{n-{K}}\left(L\right) = F_n$, similar to ideal gases. This can be understood by taking its divergence, 
 \begin{equation}
     \partial_{x}T^{xx}_{n-{K}}(x)= F_n \delta(x) - F_n \delta (x-L) \ .
 \end{equation}
The non-conserved terms correspond to the mechanical effects.  

In general, a MCD discontinuity (similar to a potential) in $D$-dimension leads to a well-defined and finite force in $(D-1)$-dimension. In the 1D well, this is a force acting on a ``point'', while in the 3D ideal gases, the discontinuity becomes a pressure exerted on a ``surface''. However, in the region where the kinetic MCD is constant, there is no force. 

In the 2D case, one observes different behaviors of the momentum flow along different directions. For example, in the rectangular box mentioned above, the kinetic MCD is, 
\begin{equation}
   T_{n_x,n_y-{K}}^{xx}=\frac{n_{x}^{2}\pi^{2}\hbar^{2}}{mL_{x}^{3}L_{y}}\sin^{2}\frac{n_{y}\pi y}{L_{y}}\theta_{\cal V}; ~~~ T_{n_x,n_y-{K}}^{yy}=\frac{n_{y}^{2}\pi^{2}\hbar^{2}}{mL_{x}L_{y}^{3}}\sin^{2}\frac{n_{x}\pi x}{L_{x}}\theta_{\cal V} \ .
\end{equation}
Thus, the wall in the $\hat{x}$-direction receives a $y$-dependent force, different from that in the $\hat{y}$-direction. This anisotropy is a characteristic of wave properties. 

The anisotropy and inhomogeneity of momentum flow can also between in a 2D circular infinite well with radius $R$. By explicitly solving the Schr\"{o}dinger's equation, one obtains the eigenstates and corresponding eigenvalues~\cite{Robinett:2003aa},
\begin{equation}
    \psi_{m,n}\left(r,\theta\right)=C_{m,n} J_{m}\left(z_{m,n}\frac{r}{R}\right)\cos (m\theta)\ ; ~~~ E_{m,n}=\frac{\hbar^{2}z_{m,n}^{2}}{2MR^{2}}\ ,
\end{equation}
where integers $m=0,1,2,...$ and $n=1,2,...$ represent energy levels, $J_m$ the Bessel function of the first kind, $z_{m,n}$ the $n$-th zero of $J_m(z)$ and $C_{m,n}$ the normalization constant. From the above, one obtains 
the kinetic MCD at the boundary ($r=R$) in the polar coordinates, 
\begin{equation}
    T^{rr}_{m,n-K}\left(r=R,\theta\right)=\frac{C_{m,n}^2 z_{m,n}^2\hbar^{2}}{2MR^2}\left[J_{m}^{\prime}\left(z_{m,n}\right)\cos \left(m\theta\right)\right]^{2} \ ,
\end{equation}
and all other components are zero $T^{r\theta}=T^{\theta r}= T^{\theta\theta}=0$. It encounters a discontinuity near the boundary, indicating a radial normal force on the wall. Moreover, for $m \neq 0$, the force has an anisotropic $\theta$ dependence and cannot simply be regarded as a pressure  in the pressure-volume work. 

Thus applicability of quantum pressure can limit possible quantum states, such as in the M.I.T. bag model~\cite{Chodos:1974je,Chodos:1974pn,Johnson:1975zp,Thomas:1982kv}, where solutions with broken spherical symmetry are ruled out by the requirement of a constant pressure on the boundary. In this model, the massless Dirac quarks are confined inside a 3D spherical cavity of radius $R$ with vanishing scalar density $\bar{\psi}\psi=0$ at the boundary. The latter can be realized with an infinite scalar potential outside~\cite{Bogolubov:1968zk}. The force on the boundary due to the discontinuity of the kinetic MCD is balanced by an external isotropic vacuum ``pressure'' $B$ (bag constant). The requirement of an isotropic MCD limits possible quark states with just total angular momentum $j=1/2$. 

In these special states, the radial component of the kinetic MCD $T^{rr}_{K}$ of one quark is independent of angular variables, 
\begin{align}
    T_{n,\kappa-K}^{rr}(\vec{r})=\frac{\omega_{n,\kappa}^{4}}{8\pi R_{0}^{4}\left(\omega_{n,\kappa}+\kappa\right)\sin^{2}\omega_{n,\kappa}} \left[j_{1}^{2}\left(t_{n,\kappa}\right)+j_{0}\left(t_{n,\kappa}\right)j_{1}^{\prime}\left(t_{n,\kappa}\right)\right] \theta_{\cal V} \hat{e}_r^i \hat{e}_r^j \ ,
\end{align}
where $j_l(t)$ is the conventional spherical Bessel function of the first kind, $\kappa=\pm1$ is the Dirac quantum number, $n$ labels frequencies determined by the boundary condition, $\omega_{n,\kappa}$ is the energy level determined by the boundary conditions and $t_{n,\kappa}=\omega_{n,\kappa}r/R$. Because of the discontinuities of the kinetic MCD near the boundary, one finds there is an isotropic 2D pressure force acting on the cavity wall from quarks, 
\begin{equation}
    p = n_\mu n_\nu T_{n,\kappa-K}^{\mu\nu}(\vec{r})\bigg{|}_{r=R} = \frac{\omega_{n,\kappa}}{4\pi R^{4}} > 0 \ .
\end{equation}
The positivity indicates a pressure outward, which is in turn balanced by the vacuum MCD $T^{ij}_{\rm B} = -B \delta^{ij} \theta_{\cal V}$, leading to a total conserved MCD everywhere, including at the boundary, 
\begin{equation}
    T^{ij}_{\rm bag}(\vec{r}) = T^{ij}_{K}(\vec{r}) + T^{ij}_B(\vec{r}); ~~~\partial_i T^{ij}_{\rm bag}(\vec{r})=0
\end{equation}
Therefore, the vacuum imposes a 2D confining force on quarks inside the cavity. Different states $n, \kappa$ are adjusted through changing the radius $R$ to yield the same outward quark pressure. 

The total MCD in the bag model, $T^{ij}_{\rm bag}(\vec{r})$, has no interpretation as a pressure force anywhere within the cavity. The individual terms can be interpreted as pressure forces only when there is a discontinuity. Note that although the angular components of the total MCD $T^{\theta\theta}$ and $T^{\varphi\varphi}$ are discontinuous along the radial directions, this type of discontinuities does not seem to have physical effects, which can be seen from Eq.~(\ref{eq:Disc-Force}). 

\subsection{3D Bound States in a Potential: H-Atom}

In more general cases, a clear boundary for quantum bound states does not exist. The kinetic MCD is continuous and inhomogeneous over space, and the kinetic pressure ceases to be a useful concept.  We then focus on the properties of the momentum flow, especially through the continuity equation, to study possible force effects. 

For particles in a potential, such as the infinite well above, their kinetic MCDs are not conserved. This non-conservation is essentially the signal for mechanical effects, reflecting that the force induces momentum changes. For eigenstates $|\varphi_n\rangle$ of a particle inside a potential $V(\vec{r})$, the continuity equation can be simplified using the Schr\"{o}dinger's equation, 
\begin{equation}
    \partial_{i}\langle \varphi_n |\hat{T}^{ij}_{K}(\vec{r})|\varphi_n\rangle = \left[-\nabla^j V(\vec{r})\right] |\varphi_n(\vec{r})|^{2} = F^j |\varphi_n(\vec{r})|^{2} \ , \label{eq:Quantum-Force}
\end{equation}
where $F^{j}$ is the classical force on a particle at $\vec{r}$. The quantum average introduces an extra factor $|\varphi_n(\vec{r})|^{2}$, which is the probability of finding a particle at $\vec{r}$. From a naive analogy to E\&M fields, this factor appears to be a ``charge'' density. However, this interpretation is inappropriate because it is not physical to view a part of the wave function as a part of the particle. Instead, $\langle T^{ij}_{K} (\vec{r}) \rangle$ is the expectation value of the momentum flow at $\vec{r}$, and its divergence is then the momentum inflow generated from the external potential, and thus a force ``density''. The total MCD then receives contributions from the potential $T^{ij}_{I}$ as well, which can be directly solved as in Eq.~(\ref{eq:MCD-Int}) for a given potential, 
\begin{equation}
    \partial_{i}\hat{T}^{ij}_{I}(\vec{r}) = \left[+\nabla^{j}V(\vec{r})\right]\delta^{(3)}(\vec{r}-\hat{\vec{r}}), ~~~ \hat{T}_{I}^{ij}(\vec{r})=\nabla^{i}\left[-\frac{\hat{\nabla}^{j}V(\hat{\vec{r}})}{4\pi|\vec{r}-\hat{\vec{r}}|}\right] \ .\label{eq:MCD-QI}
\end{equation}
And the total MCD is now conserved. 

\begin{figure}[ht]
    \centering
    \includegraphics[width=0.65\linewidth]{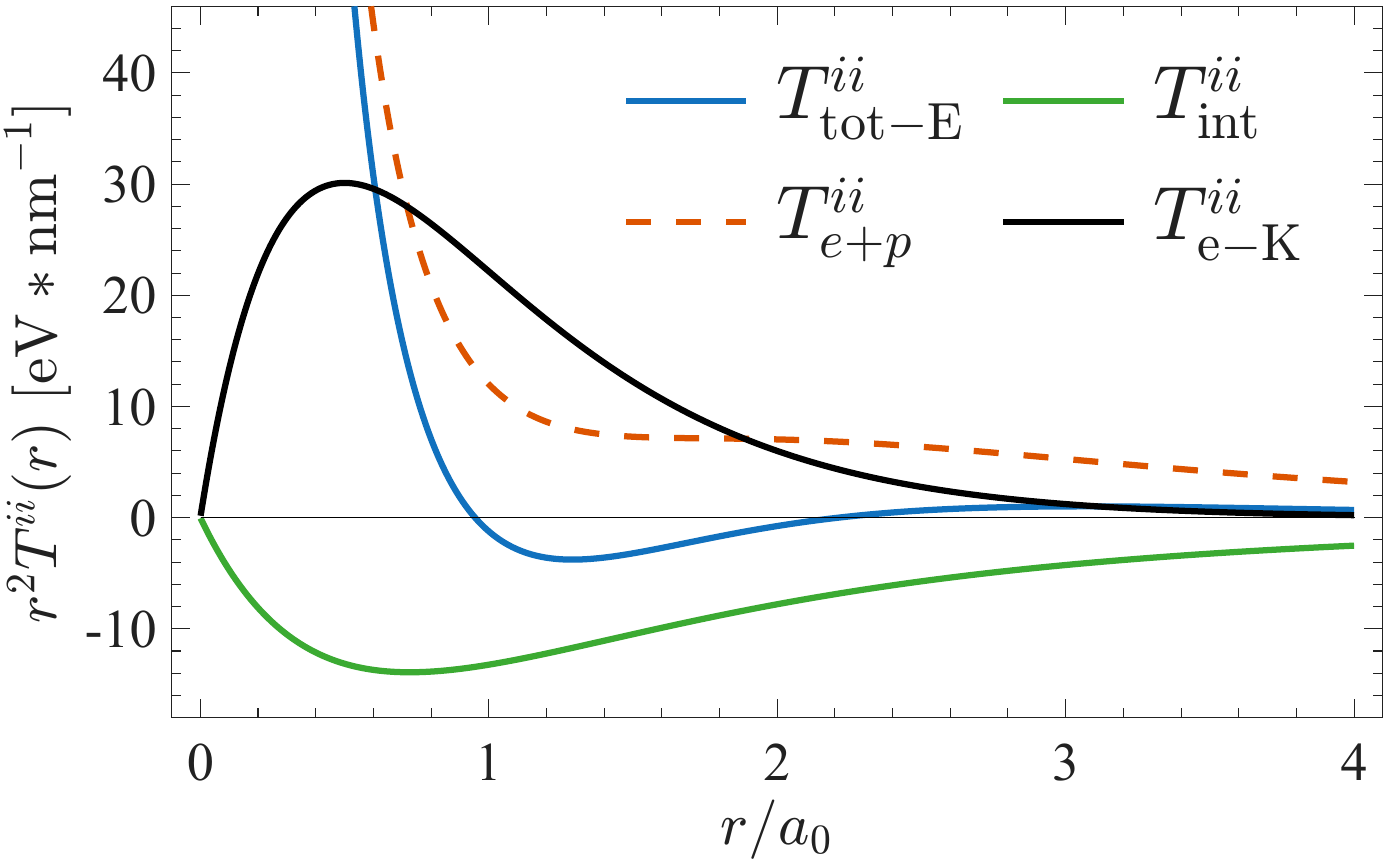}
    \caption{Traces of the momentum flows in H-atom: electron's kinetic MCD $T^{ii}_{\rm e-K}$ (black solid), interference MCD from the electron and proton fields $T^{ii}_{\rm int}$ (green solid), their self-MCDs $T^{ii}_{e+p}=T^{ij}_e+T^{ij}_p$ (brown dashed) and the total electric MCD $T^{ii}_{\rm tot-E}$ (blue solid).}
    \label{fig:H_Atom_Trace}
\end{figure}

We again consider the example of an H-atom which has been studied in details in ~\cite{Ji:2022exr, Freese:2024rkr}, where the static electric field serves as a ``messenger'' to transfer momentum between the electron and proton. For simplicity, we will consider the ground state of electron $|\psi_{1s}\rangle$ in the infinitely heavy limit of the proton. The kinetic MCD of the electron has been worked out in~\cite{Ji:2022exr}, and is positive definite, representing an anisotropic flow in angular directions, shown as the black solid line in FIG.~\ref{fig:H_Atom_Trace}. The interaction MCD from electric fields is separated into the self electric fields of the proton and electron $T^{ij}_{e}(\vec{r}), T^{ij}_{p}(\vec{r})$ and the interference part $T^{ij}_{{\rm int}}(\vec{r})$. The self-field MCDs, the interference part and the total electric MCD are also shown in FIG.~\ref{fig:H_Atom_Trace} as the brown, green and blue lines, respectively. 

The interference part is crucial for the momentum transfer~\cite{Ji:2022exr}, 
\begin{equation}
    T_{{\rm int}}^{ij}(\vec{r}) =-\frac{\hbar^{2}}{2\pi m a_{0}^5}\frac{a_0^4}{2r^{4}}\left[1-e^{-2r/a_{0}}\left(1+\frac{2r}{a_{0}}+\frac{2r^{2}}{a_{0}^{2}}\right)\right]\left(\delta^{ij}-2\frac{r^{i}r^{j}}{r^{2}}\right) \ .
\end{equation}
The trace of this tensor is negative definite. Through the momentum continuity equation, one recovers exactly the attractive Coulomb force~\cite{Freese:2024rkr,Ji:2025gsq}, 
\begin{equation}
    \partial_{i}T_{{\rm e-K}}^{ij}(\vec{r}) = - \frac{\hbar^{2}}{\pi m a_{0}^{4}r^{2}}\frac{r^{j}}{r}e^{-2r/a_{0}} = \left|\psi_{1s}(\vec{r})\right|^2 F^j(\vec{r}) = - \partial_{i}T_{{\rm int}}^{ij}(\vec{r}) \ , \label{eq:e-Force}
\end{equation}
where $F^j(\vec{r})$ is the Coulomb force from the proton acting on an electron at $\vec{r}$.

However, as we mentioned above, the total electric MCD, which can be measured through gravitational form factors, also contains the important contributions from the self electric field of the proton and electron, 
\begin{align}
    T_{e}^{ij}(\vec{r})&=\left(\delta^{ij}\nabla^{2}-\nabla^{i}\nabla^{j}\right)\left[\frac{\hbar^{2}}{32\pi^{2}ma_{0}^{4}}\int{\rm d}^{3}r_{1}e^{-2r_{1}/a}\frac{1}{\left|\vec{r}-\vec{r}_{1}\right|^{2}}\right] \ ,\\
    T_{p}^{ij}(\vec{r})&=\left(\delta^{ij}\nabla^{2}-\nabla^{i}\nabla^{j}\right)\left[\frac{\hbar^{2}}{32\pi ma_{0}r^{2}}\right] \ .
\end{align}
The above form requires special regularization of the divergent behavior near the point-like charge, which is discussed in detail in Appendix.~\ref{sec-app:point_like_charge}. It is obvious that these two MCDs are conserved themselves, and thus does not contribute to the Coulomb attractive force. However, they are crucial to cancel divergences from long-range behavior of the interference part. At large distances $r \to \infty$, the leading non-vanishing total MCD would be, 
\begin{align}
    \lim_{r\to\infty}T^{ij}
    &=\frac{\hbar^{2}}{2\pi ma_{0}^{5}}\frac{a_{0}^{6}}{r^{6}}\left(\delta^{ij}-\frac{3}{2}\frac{r^{i}r^{j}}{r^{2}}\right)\ .
\end{align}
Since the kinetic MCD of electron behaves as an exponential decay, the above total MCD comes entirely from the electric field. The $r^{-6}$ decay indicates an electric transition dipole arising from the quantum fluctuations. 

Finally, none of the interaction and kinetic MCDs have any pressure interpretation or direct force effects because anisotropic motion and long-range interaction. The traces of different contributions to the MCD, as depicted in~\cite{Ji:2025gsq}, do not represent pressure in any known way~\cite{Ji:2022exr,Czarnecki:2023yqd}. 

\subsection{Quantum Many-Body Systems}

In classical gases and liquids, homogeneity and isotropy are achieved through numerous physical degrees of freedom. If we allow many quantum particles present in a box, we might recover these features as well. Moreover, when the short-range interactions are introduced, one may define mechanical stress tensor as in continuous media. To highlight the quantum nature, we will first consider the systems at zero temperature. Finally, we will remark on the nuclear matter at finite temperature. 

\subsubsection{Quantum Bose Liquid: Pilot Wave and Pressure Force}

For a large number of identical, non-interactive bosons in a box, the ground state corresponds to all the particles occupying the lowest energy state of the single particles, resulting in a Bose-Einstein condensation. It has a spatial distribution of bosons following the profile of the probability density of single-particle wave function. Here the kinetic MCD is also inhomogeneous and anisotropic, the same as the examples in the previous subsections. When interactions are present, the system can be studied through the Gross–Pitaevskii equation under the Hartree (or single particle) approximation~\cite{rogel2013gross}.

The wave function of the many-boson system is approximated by a direct product of $N$ identical bosons as $\Psi\left(\vec{r}_{1},\vec{r}_{2},\dots\vec{r}_{N}\right)=\psi\left(\vec{r}_{1}\right)\psi\left(\vec{r}_{2}\right)\dots\psi\left(\vec{r}_{N}\right)$, where $\psi$ is the single-particle wave function satisfying the normalization condition $\left\langle \psi|\psi\right\rangle =1$. Suppose the Hamiltonian of the system is given as, 
\begin{equation}
    H=\sum_{a}\left[\frac{\vec{k}_{a}^{2}}{2m}+V\left(\vec{r}_{a}\right)\right]+\sum_{b<a}\lambda\delta^{(3)}\left(\vec{r}_{a}-\vec{r}_{b}\right)
\end{equation}
where $\vec{k}_a$ is the momentum, $V(\vec{r}_a)$ is the external potential and $\lambda$ is a parameter representing the internal {\it contact} interactions among particles. The macroscopic wave of the system is given as a rescaled single-particle wave function $\phi=N^{1/2}\psi$ with normalization $\left\langle \phi|\phi\right\rangle =N$. Therefore, one finds $\left|\phi(\vec{r})\right|^{2}=\rho(\vec{r})$ is the number density of particles, and a general form of this wave function is written as, 
\begin{equation}
\phi(r)=\sqrt{\rho(\vec{r})}e^{iS(\vec{r})/\hbar}  \ .  
\end{equation}
By examining the probability continuity equation, 
\begin{equation}
    0=\frac{\partial\rho}{\partial t}+\nabla\cdot\vec{j}=\frac{\partial\rho}{\partial t}+\nabla\cdot\left(\rho\frac{\nabla S}{m}\right)\equiv\frac{\partial\rho}{\partial t}+\nabla\cdot\left(\rho\vec{v}_{\phi}\right) \ ,
\end{equation}
where $\vec{v}_{\phi}\equiv\nabla S/m$ represents the probability current for a quantum system, and is then directly the physical velocity of the quantum bosonic liquid. 

Utilizing the kinetic MCD operator in Eq.~(\ref{eq:MCD-QuantumKinetic}), one can directly take the expectation value of this operator in the state $|\Psi\rangle$ and obtain, 
\begin{equation}
     \langle \Psi|\hat{T}_{{K}}^{ij}(\vec{r})|\Psi\rangle =\rho\left[mv_{\phi}^{i}v_{\phi}^{j}+\frac{\hbar^{2}}{2m}\left(\frac{\nabla^{i}\sqrt{\rho}}{\sqrt{\rho}}\frac{\nabla^{j}\sqrt{\rho}}{\sqrt{\rho}}-\frac{\nabla^{i}\nabla^{j}\sqrt{\rho}}{\sqrt{\rho}}\right)\right]
\end{equation}
This kinetic MCD can be separated into two parts, where the first term has similar forms to the macroscopic motions in gases and liquids (see Eqs. (\ref{eq:MCD-idealgas}) and (\ref{eq:MCD-Liquid})). The second term has been considered as a ``quantum stress tensor'' in Ref.~\cite{takabayasi_formulation_1952,Freese:2024rkr}, but is actually the effect we discussed early as ordered flow for stationary states in a box or an H-atom. In bosonic liquids, it represents kinetic MCD effects from the gradient of boson densities.

The interaction MCD can be directly solved from Eq.~(\ref{eq:MCD-QI}), and the expectation value gives the force density acting on the system, 
\begin{equation}
    \langle \Psi|\partial_{i}\hat{T}_{I}^{ij}(\vec{r})|\Psi\rangle =\nabla^{j}\left[\frac{1}{2}\lambda\rho^{2}(\vec{r})\right]+\rho(\vec{r})\nabla^{j}V(\vec{r})\equiv\nabla^{j}p_{I}(\vec{r})+\rho(\vec{r})\nabla^{j}V(\vec{r})
\end{equation}
The first term arises from the zero-range internal interactions among particles, which can be effectively taken as a macroscopic contact pressure within the bosonic liquid. However, since the interaction MCD receives ambiguities from the superpotentials, one must be careful in extracting such contact forces from that. For example, ignoring the external potential $V(\vec{r})$, the solution to the interaction MCD as in Eq.~(\ref{eq:MCD-QI}) is given as, 
\begin{equation}
    \langle \Psi|\hat{T}_{I-{\rm in}}^{ij}(\vec{r})|\Psi\rangle = p_I(\vec{r})\delta^{ij}+\left(\delta^{ij}\nabla^{2}-\nabla^{i}\nabla^{j}\right)\int{\rm d}^{3}\vec{x} \frac{p_I(\vec{x})}{4\pi\left|\vec{r}-\vec{x}\right|}
\end{equation}
In the previous examples of interaction pressure in gases and stress tensor in liquids and solid, one finds the boundary conditions play a crucial role in determining the physical meaning of the interaction MCD. The first term of the above solution obviously satisfies the same boundary conditions where the MCD vanishes in the vacuum. Therefore, one may directly identify $p_I$ as the physical contact force within such bosonic liquid. This result is consistent with that from the pressure-volume work of statistical mechanics~\cite{pathria2017statistical}, as well as the interaction pressure in classical gases. 

The divergence of the kinetic MCD can be written as,
\begin{equation}
    \mathcal{F}^j = \langle \Psi|\partial_{i}\hat{T}_{{K}}^{ij}(\vec{r})|\Psi\rangle =\rho\left[m\left(\vec{v}_{\phi}\cdot\nabla\right)v_{\phi}^{j}-\frac{\hbar^{2}}{2m}\nabla^{j}\left(\frac{\nabla^{2}\sqrt{\rho}}{\sqrt{\rho}}\right)\right]
\end{equation}
This originates from the force density acting on the local particles of the bosonic liquid, which is the divergence ofthe interaction MCD. The balance equation or the momentum continuity equation is now exactly the Gross–Pitaevskii equation, 
\begin{equation}
    \partial_{i}\langle\hat{T}_{{K}}^{ij}(\vec{r})+\hat{T}_{I}^{ij}(\vec{r})\rangle \propto m(\vec{v}_{\phi}\cdot\nabla)v_{\phi}^{j}-\nabla^{j}\left[\frac{\hbar^{2}}{2m}\frac{\nabla^{2}\sqrt{\rho}}{\sqrt{\rho}}-V(\vec{r})-\lambda\rho\right] = 0 \ .
\end{equation}
Notably, these formulas are formally very similar to a recent application of pilot wave interpretation to the H-atom~\cite{Freese:2024rkr}. It is interesting that the formalism is physically realizable in a quantum many-body system in the classical limit. 

It might be useful to think of the Skyrme model in the above framework, in which the interaction MCD may be interpreted as surface forces in a pion liquid. However, even in the large $N_c$ limit of QCD, this might not be true in the nucleon due to the absence of short-range forces~\cite{Witten:1979kh} between quarks. In the case of spontaneous symmetry breaking of electroweak symmetry, the ground state may also be considered as a boson condensate. 

Before ending our discussion on many-boson systems, we remark that light waves can be considered as many photons in the same quantum state, with classical fields corresponding to the quantum expectations in a coherent state. At finite temperature, one of the well-known examples of many-boson system is the black-body radiation. From the statistical mechanics viewpoint, the system is similar to an ideal gas with Bose-Einstein statistics. 

\subsubsection{Degenerate Fermi Gas: Kinetic Pressure and Long-Range Force}\label{sec:fermi}

If the constituent particles are identical fermions with mass $m$, the Pauli exclusion principle requires all the single-particle energy levels below the Fermi energy $\varepsilon_F$ to be fully occupied. When the number of fermions $N$ becomes very large, the discrete feature of energy level $\varepsilon$ or 3-momentum $k$, which depends on the geometry of the box, becomes less important. In the non-relativistic limit, the Fermi momentum $k_F$ and Fermi energy $\varepsilon_F$ are well known, 
\begin{equation}
    k_{F} = 2\pi\hbar\left(\frac{3N}{4\pi gV}\right)^{1/3}\ ; ~~~ \varepsilon_F = \frac{k_F^2}{2m}\ ,
\end{equation}
where $g$ is the degeneracy factor. Due to the presence of many particles occupying a large number of quantum states in the momentum space, homogeneity and isotropy in coordinate space is recovered. Therefore, one obtains the kinetic MCD in a similar way for classical ideal gases, which again represents the average kinetic energy of fermions, 
\begin{equation}
    \langle \hat{T}^{ij}_{K} \rangle = \frac{2}{3} n \varepsilon_K \delta^{ij} \theta_{\cal V}\ ; ~~~ \varepsilon_K = \frac{3}{5}\varepsilon_ F \ ,\label{eq:MCD-Fermi}
\end{equation}
where $n=N/V$ is the number density and $\varepsilon_K$ is the average kinetic energy of one fermion. 

This MCD gives directly the pressure through its discontinuities near the boundary, similar to the ideal gases, 
\begin{equation}
    p_{K} = \left[ \langle \hat{T}^{ij}_{K} \rangle_{\rm in}- \langle \hat{T}^{ij}_{K} \rangle_{\rm out}\right] n^i n^j = \frac{2}{5} n \varepsilon_F \ .
\end{equation}
This result is the same as the one obtained from quantum statistical mechanics through the pressure-volume work. 

When interactions are present and the particle density is high, the system may satisfy the scale separation and reach local equilibrium as in a classical fluid, where each element can be seen as an ideal Fermi gas with local density $n(\vec{r})$ and kinetic pressure $p_K(\vec{r})$. This model has been used for a large-$Z$ atom in the Thomas-Fermi model~\cite{thomas1927calculation,fermi1927statistical}, neutron stars~\cite{Strobel:1997vf}, and many other similar systems.

The total MCD in the Thomas-Fermi model contains two parts: the kinetic motion in a local gas element and the long-range Coulomb interactions,
\begin{equation}
   T^{ij}(\vec{r}) = \delta^{ij} p_{K}(\vec{r}) + T^{ij}_{I}(\vec{r}) \ ,
\end{equation}
where the kinetic pressure $p_K(\vec{r})$ depends on the local density. In a large-$Z$ atom, the local density of electrons is approximately proportional to $Z$ as the size of an atom is roughly $Z$-independent. Therefore, the local Fermi gas approximation may be suitable. In this picture, an electron experiences an attractive field $\vec{E}_Z$ from the positive nucleus $Ze$ at the center as well as the repulsive field $\vec{E}_e$ from other electrons. Consequently, the interaction MCD from Coulomb interactions can be written as, 
\begin{equation}
    T^{ij}_{I} =  \frac{1}{2} \epsilon_0(\vec{E}_e+\vec{E}_Z)^2 \delta^{ij} -\epsilon_0(E^i_e+E^i_Z)(E^j_e+E^j_Z) \ .
\end{equation}
In principle, the above shall be considered as an expectation in the Fermi gas state. At high density, the quantum correlation (Fock term) can be neglected, and one can directly solve the classical electric field (Hartree term) from Gauss's Law, 
\begin{equation}
    {\nabla} \cdot ( \vec{E}_e+\vec{E}_Z ) = -\nabla^2 \Phi(\vec{r}) = \frac{1}{\epsilon_0} \left[ -en(\vec{r})+Ze\delta^{(3)}(\vec{r}) \right] \ ,
\end{equation}
where $\Phi(\vec{r})$ is the local electric potential. One may further assume that the system is spherically symmetric, and the hydrostatic equation can be derived from the momentum continuity equation, $\partial_i T^{ij} =0$, which reduces to, 
\begin{equation}
    \nabla p_{K} (\vec{r}) + en(\vec{r}) (\vec{E}_e+\vec{E}_Z) = \nabla p_{K} (\vec{r}) - en(\vec{r}) \nabla\Phi = 0 \ . \label{eq:Thomas}
\end{equation}
This equation is the famous Thomas-Fermi equation~\cite{thomas1927calculation,fermi1927statistical}. Combined with the relation between the local pressure and density, the above equations can be solved self-consistently. 

The pressure gradient in Eq.~(\ref{eq:Thomas}) appears similar to that in the Euler's equation of fluid mechanics, where it does represent a contact force. However, the degenerate pressure arises from the kinetic motion of a large number of electrons, different from the interaction pressure in liquids, and therefore does not represent any surface forces. The physics of the above equation is that the kinetic momentum flow gradient is balanced by static electric force, similar to the model of the earth's atmosphere. Consequently, no direct mechanical interpretation such as surface pressure force can be assigned to any part of the total MCD. 

\subsubsection{Nuclear Matter}\label{sec:nuclmattter}

Nuclear matter is a system of interacting protons and neutrons, extensively studied in the field of nuclear physics. Ideally, it extends uniformly without Coulomb (long-range) or surface effects, serving as the fundamental reference for both finite nuclei and the neutron stars. Through studying the equation of state and relating quantities such as energy density and pressure in thermodynamics, researchers have obtained a profound understanding of the nuclear and stellar structure~\cite{Moller:1993ed,Li:2008gp}. 

For a uniform nuclear matter system consisting of $N$ nucleons in a box of volume $\cal V$, the isotropic MCD is $T^{ij}= p \theta_{\cal V} \delta^{ij}$, where $p$ is the pressure---the discontinuity of MCD across the boundary. One can in principle obtain the unique pressure $p=p_K+p_I$ by solving the momentum continuity equation with the boundary condition that $p_I$ vanishes outside the box. However, as in the case of ideal gas, one can also obtain this pressure simply by taking the derivative of the ground state energy with respect to the volume, $p=-{\rm d}E/{\rm d}{\cal V}$, which is the standard practice in the literature~\cite{Mosel:1974ntk,Sauer:1976zzf,Li:2008gp}. 

While the kinetic pressure $p_K$ is always positive, the interaction part $p_I$ depends on the density of the nuclear matter. In highly dense nuclear matter where the repulsive part of the hard-core potential dominates, the interaction pressure $p_I$ is positive. Moreover, at very high density, such as at the center of a neutron star, $p_I$ completely dominates over $p_K$, and one may consider the system is a continuous medium with the pressure approximates the contact force pushing against different parts of the system. 

However, in dilute nuclear matter, $p_I$ becomes negative due to attractive part of the nuclear forces. As the density decreases and at some point, $p_I$ finally cancels $p_K$, resulting in zero total pressure, and we obtain the nuclear matter at the center of a large nucleus. Note that even through the total pressure vanishes, there are attractive interactions which generate a negative pressure. If the volume further expands, one no longer have a uniform nuclear matter system as the attractive interactions will keep the nucleons together to lower the total energy. For a finite nucleus, the central pressure is slightly positive $p_K>|p_I|$ due to the surface tension, but the surface contact force $p_I$ is still attractive. 

\section{Physics of MCD in QCD and Color-Lorentz Forces in the Nucleon}\label{sec:MCD-QCD}

We finally come to investigating the central topic of this paper: physics of the MCD within the nucleon, specifically the proton. By employing a well-established decomposition of the QCD EMT~\cite{Ji:1994av}, we identify different contributions to the QCD MCD from quark kinetic term, gluon tensor term and gluon scalar term (trace anomaly). Unlike the electron momentum flow in the H-atom purely in the angular directions, quarks inside the proton also exhibit a radial kinetic flow component. There is also a negative contribution to the momentum flow from the QCD trace anomaly. 

The actual mechanical effect on quarks inside the proton is related to the divergence of its kinetic MCD. By using phenomenological form factors, one observes an attractive, inward-directed color-Lorentz force, indicating the confinement effect. This attractive force can be decomposed into individual contributions from the gluon tensor and scalar components. While the gluon tensor contributes a repulsive force---reflecting the dominance of radiative gluons---the trace anomaly is the origin of the strong confining force. More interestingly, the average magnitude of this anomaly force is approximately 1 GeV/fm, closely matching the well-known QCD string tension. 

\subsection{QCD Energy-Momentum Tensor, Renormalization and Form Factors}

In this subsection, we review the definition, renormalization and form factors of the QCD EMT. The form that has been mostly used is the gauge-invariant and symmetric second order tensor, and is believed to be the gravitational charge~\cite{Jaffe:1989jz,Ji:1994av}. Due to ultraviolet (UV) behaviors, the individual terms are divergent and their renormalization properties have been known since 1970's and now is a textbook material~\cite{Peskin:1995ev}. 

More specifically, this operator can be split into trace $\hat{T}^{\mu\nu}$ and traceless $\bar{T}^{\mu\nu}$ parts~\cite{Ji:1994av} 
\begin{equation}
    T^{\mu\nu}_{\rm QCD} = \bar{T}^{\mu\nu} + \hat T^{\mu\nu}
\end{equation} 
where both parts are scale and scheme-independent. The traceless part can be written as a sum of renormalized quark and gluon operators,  $\bar{T}^{\mu\nu}=\bar{T}^{\mu\nu}_q(\mu)+\bar{T}^{\mu\nu}_g(\mu)$, which now depend on the renormalization scheme and scale $\mu$, and enter in the operator product expansion for the inclusive deep-inelastic scattering. Obviously the scale and scheme dependence cancels in the sum. The renormalization mixing of the quark and gluon parts was first studied in Refs.~\cite{Gross:1973zrg,Georgi:1974wnj} and now has been calculated to 4-loops~\cite{Moch:2021qrk}. The anomalous contribution (trace anomaly) to the trace part $\hat{T}^{\mu\nu}$ was calculated in Ref.~\cite{Collins:1976yq} in dimensional regularization. In lattice regularization, the trace anomaly has been calculated in perturbation theory at one-loop level~\cite{Caracciolo:1989pt,Caracciolo:1991cp}. A non-perturbative derivation of the trace anomaly on lattice for heavy-quark potential and its contribution to the energy sum rule has also been carried out~\cite{Rothe:1995hu,Rothe:1995av}. Of course, the trace anomaly must be independent of renormalization scheme and scale. A general discussion about the trace anomaly contribution to the QCD Hamiltonian has been made in Ref.~\cite{Ji:2021qgo}. 

Therefore, the QCD EMT naturally receives three (renormalized) contributions, 
\begin{align}
    \bar{T}_{q}^{\mu\nu}(\mu) = & \frac{1}{2}\bar{\psi}i\overleftrightarrow{\mathcal{D}}^{(\mu}\gamma^{\nu)}\psi-\frac{1}{4}g^{\mu\nu}\bar{\psi}m\psi \ ,  \nonumber \\
    \bar{T}_{g}^{\mu\nu}(\mu)= & \frac{1}{4}g^{\mu\nu}F^{2}-F^{\mu\alpha}F_{\ \alpha}^{\nu} \ , \\
    \hat{T}^{\mu\nu}= & \frac{1}{4}g^{\mu\nu}\left[\left(1+\gamma_{m}\right)\bar{\psi}m\psi+\frac{\beta\left(g\right)}{2g}F^{2}\right]\nonumber \ ,  
\end{align}
where we have ignored operators that have vanishing contributions in the physical states when a (covariant) gauge fixing is needed~\cite{Ji:1995sv}. If we further split the trace operator, the most natural idea is to consider the quark mass operator $\bar \psi\psi$ and the gluon scalar $F^2$ as separate quark and gluon contributions~\cite{Ji:1994av}. 

There has been other attempts in the literature to split the trace part by starting from the bare quark and gluon parts of EMT $T^{\mu\nu}_{\rm QCD} =T^{\mu\nu}_{0q}+T^{\mu\nu}_{0g}$ and performing renormalization to get renormalized quark and gluon operators, $T^{\mu\nu}_{R q,g}$. An example in dimensional regularization has been worked out by Hatta and Tanaka~\cite{Hatta:2018sqd}. The finite traces arise separately from divergent bare quark and gluon tensors and have been designated accordingly as quark and gluon anomaly contributions. On the other hand, the anomaly can arise from different forms of UV violations of the classical scale symmetry. For instance, in the case of the axial anomaly, the result can arise from the integration measure of the path integral whereas both current and action are invariant under axial symmetry~\cite{Peskin:1995ev}. Focusing on the UV aspect of the EMT operator can lead to the incorrect conclusion that the anomaly does not exist in the renormalized Hamiltonian density, $T^{00}_{\rm QCD}$~\cite{Metz:2020vxd,Lorce:2021xku}. 

As clear from previous sections, there are two main categories of MCDs, the kinetic motion of matter particles, quarks in this case, and the interaction effect of gluon fields. The traceless quark MCD $\bar{T}^{ij}_q$ contains two terms, the kinetic motion of quarks $\bar{\psi} \mathcal{D}^{(i} \gamma^{j)} \psi$, and a scale-independent scalar operator $\bar{\psi}m\psi$. In the previous study~\cite{Ji:2025gsq}, we have ignored the light quark mass effects which we intend to keep in the following discussions, and of course, $\bar{\psi}m\psi$ is also part of the trace. The gluon tensor includes static Coulomb fields and radiative (off-shell) gluons. Since there is no gauge-invariant and frame-independent separation between them, the same as in QED, these contributions shall be collectively considered, denoted as $\bar T_g^{ij}$. Therefore, we are led to a natural rearrangement, 
\begin{align}
    T_{q}^{ij} & =\frac{1}{2}\bar{\psi}i\overleftrightarrow{\mathcal{D}}^{(i}\gamma^{j)}\psi \ ,  \nonumber \\
    \bar{T}_{g}^{ij} & =-\frac{1}{4}\delta^{ij}F^{2}-F^{i\alpha}F_{\ \alpha}^{j} \label{eq:EMT-QCD} \ , \\
    T_{a}^{ij} & =-\frac{1}{4}\delta^{ij}\left[\gamma_{m}\bar{\psi}m\psi+\frac{\beta\left(g\right)}{2g}F^{2}\right]  \ . \nonumber 
\end{align}
For simplicity, our analysis below ignores anomalous dimension of the quark mass term, and the anomaly is a pure gluon scalar term. This approximation will not affect our main conclusions ($\gamma_m = 0.295$ in $\overline{\rm MS}$ at $\mu = 2$~GeV). 

The matrix elements of the above operators in the nucleon can be described in terms of the gravitational form factors (GFFs) $A_{q,g}(q^2)$, $B_{q,g}(q^2)$, $C_{q,g}(q^2)$, where $q=P'-P$ is the four-momentum transfer between the initial and final states with momenta $P$ and $P'$, respectively~\cite{Kobzarev:1962wt,Pagels:1966zza,Ji:1996ek}. Indeed it is easy to work out, 
\begin{align}
    \left\langle P^{\prime}\left|T_{q}^{\mu\nu}\right|P\right\rangle =\bar{U}\left(P^{\prime}\right)&\left[T_{q,\perp}^{\mu\nu}\left(\bar{P},q\right)-\frac{1}{4}g^{\mu\nu}M\left(G_{s,q}(q^2)-F_{\sigma}(q^2)\right)\right]U\left(P\right) \label{eq:FF-q}\\
    \left\langle P^{\prime}\left|\bar{T}_{g}^{\mu\nu}\right|P\right\rangle =\bar{U}\left(P^{\prime}\right)&\left[T_{g,\perp}^{\mu\nu}\left(\bar{P},q\right)-\frac{1}{4}g^{\mu\nu}MG_{s,g}(q^2)\right]U\left(P\right)\label{eq:FF-g}\\
    \left\langle P^{\prime}\left|T_{a}^{\mu\nu}\right|P\right\rangle =\bar{U}\left(P^{\prime}\right)&\left[\frac{1}{4}g^{\mu\nu}M\left(G_{s}(q^2)-F_{\sigma}(q^2)\right)\right]U\left(P\right) \label{eq:FF-tr}
\end{align}
where $\bar{P}^{\mu}=\left(P^{\prime\mu}+P^{\mu}\right)/2$, $\sigma^{\mu\nu}=\frac{i}{2}\left[\gamma^{\mu},\gamma^{\nu}\right]$, and the spinor normalization is given by $\overline{U}(P)U(P)=2E_P$ with $E_P$ being the energy. $F_\sigma(q^2)$ is called the sigma form factor for the term $\bar{\psi} m \psi$, $T^{\mu\nu}_{i\perp}$ is the conserved part of the EMT (orthogonal to $q^\mu$) and $G_{s,i}(q^2)$ is the scalar form factor~\cite{Ji:2021mtz} which can be written in terms of $A_{q,g},B_{q,g},C_{q,g}$ as, 
\begin{align}
    T_{i,\perp}^{\mu\nu}\left(\bar{P},q\right)&=A_{i}(q^2)\gamma^{(\mu}\bar{P}^{\nu)}+B_{i}(q^2)\frac{\bar{P}^{(\mu}i\sigma^{\nu)\alpha}q_{\alpha}}{2M}+C_{i}(q^2)\frac{q^{\mu}q^{\nu}-q^{2}g^{\mu\nu}}{M}\\
    G_{s,i}(q^2)&=A_{i}(q^2)+B_{i}(q^2)\frac{q^{2}}{4M^{2}}-C_{i}(q^2)\frac{3q^{2}}{M^{2}} \label{eq:sFF}
\end{align}
with $i=q,g$ and the total scalar form factor is $G_s = G_{s,q}+G_{s,g}$.  

\subsection{Local Spatial Densities in Breit Frame and Large $N_c$ Limit}

To visualize the structure of the nucleon, it has been a common practice to Fourier-transform the form factors into the coordinate space~\cite{Sachs:1962zzc}. While this is legitimate for non-relativistic systems where the recoils are small, it is problematic for the nucleon when the momentum transfer is comparable to the nucleon mass. This issue has been a continuing debate for many years, and a recent analysis can be found in Ref.~\cite{Jaffe:2020ebz}. 

In extracting the spatial distribution from measuring the matrix elements, there are three scales involved---the Compton wavelength related to particles' masses as $\lambda \sim 1/M$, the intrinsic scale of the spatial distribution $R$ as well as the measurement resolution $\Delta_R$. According to the uncertainty principle, $\Delta_R$ is inversely proportional to the momentum spread in the wave-packet which localizes the particle.  With these three scales, one finds in the limit $R^2 \gg \Delta_R^2 \gg \lambda^2$, the expectation value of the EMT in a localized hadron state reduces to the Fourier transformation of the matrix element in momentum space in the so-called Breit frame to be discussed later, 
\begin{equation}
    \left\langle \Psi\left|T^{\mu\nu}(\vec{r})\right|\Psi\right\rangle =\int\frac{{\rm d}^{3}\vec{q}}{\left(2\pi\right)^{3}2E_P}e^{-i\vec{q}\cdot\vec{r}}\left\langle P^{\prime}\left|T^{\mu\nu}\right|P\right\rangle
\end{equation}
The physical meaning of this limit is clear. The first limit $R^2 \gg \Delta_R^2$ means the resolution of measurement must be much smaller than the characteristic length of the internal structure (such as the charge distribution). Meanwhile, $\Delta_R^2 \gg 1/M^2$ ensures that the resolution is much larger than the Compton wavelength of the hadron, so that the measurement and internal structures will not be smeared by the quantum fluctuations. 

For the nucleon with radius $R\sim $ 1~fm, and mass $\sim 1$~GeV or $\lambda \sim $ 0.2~fm, there is little room to establish the above hierarchy. As summarized in a recent paper~\cite{Miller:2025zte}, it is impossible to obtain a rigorous 3D time-independent density from any experimental momentum-dependent form factor. One option is to construct a wave-packet of zero-momentum (delta function distribution) and infinite spatial size~\cite{Ernst:1960zza,Sachs:1962zzc}, which leads to Sachs's densities. However, this limit $\Delta_R\gg R,\lambda$ has been criticized recently in~\cite{Miller:2018ybm,Miller:2025zte}. Another approach is to first take the limit $\Delta_R \to 0$ in~\cite{Epelbaum:2022fjc,Panteleeva:2024vdw}. Under this limit, the localization uncertainty is much smaller than the Compton wavelength of the hadron. The authors found that the dependence on the particle mass is absent. However, there are two issues in this approach: first, the wave packet is dominated by large momentum components, and the resulting distribution receives a significant Lorentz boost effect. Second, any force used to localize the particle or form the wave-packet will create particle-antiparticle pairs. 

It has also been suggested to seek spatial pictures of the nucleon in IMF~\cite{Burkardt:2002hr,Miller:2007uy,Freese:2021czn}, where the infinitely large longitudinal momentum allows the localization of the nucleon or hadrons of any mass in the transverse direction. This is because the effective Compton wavelength in the transverse direction is zero, and one can define 2D transverse space density distributions. In a recent paper~\cite{Miller:2025zte}, 2D density interpretations of the axial-vector form factor and all three GFFs are obtained, along with a 2D mass density related to the trace of the EMT. We shall point out, however, that the higher-twist quantities, such as the transverse-space components of MCD, do not have simple partonic interpretation even though they are densities in the IMF. 

One might consider a limit of QCD, in which the scale separation discussed by Jaffe and hence proper densities can be obtained. However, this cannot be done simply by taking the infinite mass limit. An appropriate approach to keep a hadron radius finite while taking its mass to infinity is the large $N_c$ limit. Since there is only one scale, $\Lambda_{\rm QCD}$, in the chiral limit of QCD, the hadron radius and mass must obey the following power counting law~\cite{tHooft:1973alw,Witten:1979kh}, 
\begin{equation}
    R \sim \frac{1}{\Lambda_{\rm QCD}}= \mathcal{O}\left(N_c^0\right), ~~~ M\sim N_c \Lambda_{\rm QCD} = \mathcal{O}\left(N_c^1\right) \ . 
\end{equation}
The GFFs, on the other hand, follow different power counting laws which can be obtained from the analysis of generalized parton distributions in the large $N_c$ limit within the context of the chiral quark-soliton model~\cite{Goeke:2001tz,Goeke:2007fp},
\begin{equation}
    A(q^2) = \mathcal{O}\left(N_c^0\right), ~~~B(q^2) = \mathcal{O}\left(N_c^0\right), ~~~C(q^2) = \mathcal{O}\left(N_c^2\right)
\end{equation}
Notably, it has been proposed in~\cite{Kharzeev:2021qkd} to define the mass radius of the nucleon using the scalar form factor in Eq.~(\ref{eq:sFF}) in the infinitely heavy hadron limit. However, the mass form factor is given as~\cite{Goeke:2007fp, Cebulla:2007ei, Ji:2021mtz}, 
\begin{align}
    G_{m}(q^2) = A(q^2) + B(q^2)\frac{q^2}{4M^2} - C(q^2)\frac{q^2}{M^2} = G_s(q^2) + C(q^2)\frac{2q^2}{M^2}
\end{align}
Therefore, one finds that the difference between mass and scalar form factors does not vanish at the large $N_c$ limit, rather, it is finite of the same order $\mathcal{O}(N_c^0)$ as $G_s$ and $G_m$. Therefore, the scalar and mass form factors are different quantities and represent different aspects of the hadron structure. 

With the finite nucleon mass, to obtain 3D pictures, the E\&M properties of the nucleon such as charge distribution have been studied using Sachs's formalism and the Breit frame~\cite{Sachs:1962zzc,Jaffe:2020ebz}. Here we take the same approach for the GFFs to maintain 3D information by ignoring the subtleties about the nucleon localization, following Polyakov et al.~\cite{Polyakov:2002wz,Polyakov:2002yz}. In this frame, the initial and final momenta of the nucleon are $P=p-\frac{1}{2}q,P^{\prime}=p+\frac{1}{2}q$ such that $p^{\mu}=(E,0,0,0)$ and $q^{\mu} = (0,\vec{q})$. Therefore, the above matrix element of the MCD can be expressed as, 
\begin{align}
    \left\langle T_{q}^{ij} \right\rangle (\vec{q}) &= +\frac{M}{4}\left[G_{s,q}(\vec{q}^2) - F_\sigma(\vec{q}^2)\right]\delta^{ij}+C_{q}(\vec{q}^2)\frac{q^{i}q^{j}-\delta^{ij}\vec{q}^2}{M} \ , \\
    \left\langle \bar{T}_{g}^{ij} \right\rangle (\vec{q}) &= +\frac{M}{4}G_{s,g}(\vec{q}^2)\delta^{ij}+C_{g}(\vec{q}^2)\frac{q^{i}q^{j}-\delta^{ij}\vec{q}^2}{M} \ , \\
    \left\langle T_{a}^{ij} \right\rangle (\vec{q})&=-\frac{M}{4}\left[G_{s}(\vec{q}^2) - F_\sigma(\vec{q}^2) \right]\delta^{ij} \ .
\end{align}
For simplicity, we denote $\left\langle T^{ij} \right\rangle = \left\langle P^{\prime}\left|T^{ij}\right|P\right\rangle/2E_P$ and ignore the scale dependence in the form factors due to renormalization. In going to the coordinate space, one performs the Fourier transformation, 
\begin{align}
    T^{ij}(\vec{r}) &= \int\frac{{\rm d}^{3}\vec{q}}{\left(2\pi\right)^{3}}e^{-i\vec{q}\cdot\vec{r}}\left\langle T^{ij} \right\rangle (\vec{q}) \ , \\ 
    F(\vec{r}) &= \int\frac{{\rm d}^{3}\vec{q}}{\left(2\pi\right)^{3}}e^{-i\vec{q}\cdot\vec{r}}F(\vec{q}^2) \ , 
\end{align}
where the form factor $F$ can be $A_{q,g},C_{q,g}$ and $F_{\sigma}$.

\subsection{Phenomenological MCDs in the Nucleon}

The GFFs play a crucial role in extracting mechanical structures of the hadrons. Recently, there have been a lot of progress in studying these form factors from both theoretical and experimental viewpoint. Key advances have been made in identifying experimental processes to study them through deeply virtual Compton scattering process~\cite{Ji:1996nm,Ji:1996ek} and $J/\psi$ photoproduction process~\cite{Hatta:2018ina,Hatta:2018sqd,Hatta:2019lxo,Mamo:2019mka,Guo:2021ibg,Mamo:2022eui}, providing precise measurements~\cite{Burkert:2018bqq,Kumericki:2019ddg,Duran:2022xag}. On the theoretical front, there have been the proposal of various models, including the bag model~\cite{Ji:1997gm}, Skyrme model and chiral quark soliton model in large $N_c$ limit~\cite{Cebulla:2007ei,Goeke:2007fq,Goeke:2007fp}, chiral perturbation theory~\cite{Donoghue:1991qv}, dispersion relations~\cite{Pasquini:2014vua,Broniowski:2024oyk,Cao:2024zlf} and lattice QCD calculations~\cite{Shanahan:2018pib,Shanahan:2018nnv,Pefkou:2021fni,Hackett:2023nkr,Hackett:2023rif}. Additionally, a recent study~\cite{Stegeman:2025sca} characterizes the trace anomaly as a manifestation of spontaneous scale symmetry breaking---identifying the dilaton as the associated Goldstone boson---which provides new perspectives on GFFs and relates the dilaton to the $\sigma$ meson. 

To demonstrate the momentum current decomposition, we employ a recent global analysis on both lattice and experimental data~\cite{Guo:2025jiz}, which provides the most comprehensive and realistic information currently available. To be specific, the GFFs $F_a(q^2)$ are parametrized using the multipole ansatz with two parameters $F_a(0)$ and $M_{F_a}$ as, 
\begin{equation}
    F_a(q^2) = \frac{F_a(0)}{\left( 1-q^2/M_{F_a}^2 \right)^{\alpha}} \ , \label{eq:GFFs-Parameter}
\end{equation}
where we choose $\alpha=2$ (dipole) for $A_{q,g}$ form factors and $\alpha=3$ (tripole) for $C_{q,g}$ form factors. Notably, the $B$-form factor is found to be significantly small in recent analysis~\cite{Pefkou:2021fni,Hackett:2023rif}, and therefore shall be neglected. Since these parameters are correlated, one cannot simply add the errors in quadrature. Instead, we utilize $2\times 10^4$ parameter sets obtained from the Monte-Carlo Markov Chain (MCMC) sampling method~\cite{Guo:2025jiz}, which include the full parameter correlations. Meanwhile, the sigma form factor $F_{\sigma}(q^2)$ is also parameterized using the above dipole Ansatz, with parameters obtained using the chiral quark soliton model in the large-$N_c$ limit~\cite{Schweitzer:2003sb}. As we will see below, this sigma form factor, proportional to the quark mass, is found rather small, consistent with previous findings~\cite{Yang:2015uis,Fujii:2025aip}. 

With the Breit-frame Fourier-transformed coordinate space and the above phenomenological analysis, one can obtain the static MCDs in terms of the GFFs as, 
\begin{align}
    \langle T_{q}^{ij} \rangle(\vec{r})&=+\frac{M}{4}\left[G_{s,q}(r)-F_{\sigma}(r)\right]\delta^{ij}+\frac{1}{M}\left(\delta^{ij}\nabla^{2}-\nabla^{i}\nabla^{j}\right)C_{q}(r)\\
    \langle T_{g}^{ij} \rangle(\vec{r})&=+\frac{M}{4}G_{s,g}(r)\delta^{ij}+\frac{1}{M}\left(\delta^{ij}\nabla^{2}-\nabla^{i}\nabla^{j}\right)C_{g}(r)\\
    \langle T_{a}^{ij} \rangle(\vec{r})&=-\frac{M}{4}\left[G_{s}(r)-F_{\sigma}(r)\right]\delta^{ij} \equiv p_a(r)\delta^{ij} 
\end{align}
The above expression for MCDs in terms of isotropic and divergence-free contributions is based on the spherical symmetry of the static nucleon. By summing up the above components, one obtains the total MCD, 
\begin{equation}
    \sum_{\alpha}\langle T_{\alpha}^{ij} \rangle (\vec{r}) = -\frac{1}{M}(\nabla^i \nabla^j -\delta^{ij}\nabla^2)C(r) \ .
\end{equation}
which is a divergence-free term related to the $C/D$ form factor. Using the energy densities of each component defined by $\varepsilon_{q,g,a}\equiv T^{00}_{q,g,a}$, we can furthermore express the von Laue condition for the trace of total MCD $\langle T^{ii}\rangle$ as
\begin{equation}
    \int \langle T^{ii}(\vec{r})\rangle~{\rm d}^3\vec{r} = \int\left[\varepsilon_q(\vec{r})+\varepsilon_g(\vec{r})-3\varepsilon_a(\vec{r})\right]{\rm d}^3\vec{r}=0 \ , \label{eq:Virial}
\end{equation}
where the expression $\langle T^{ii}\rangle(\vec{r}) =\varepsilon_q(\vec{r})+\varepsilon_g(\vec{r})-3\varepsilon_a(\vec{r})$ indicates the physical significance encoded in this total momentum flow. 

By propagating the MCMC samples through the above MCD formulas---which preserves the full parameter correlations---we show the medians as curves and the uncertainties, which are estimated using 90\% confidence intervals, as the shaded areas.

\begin{figure}[ht]
    \centering
    \includegraphics[width=0.65\linewidth]{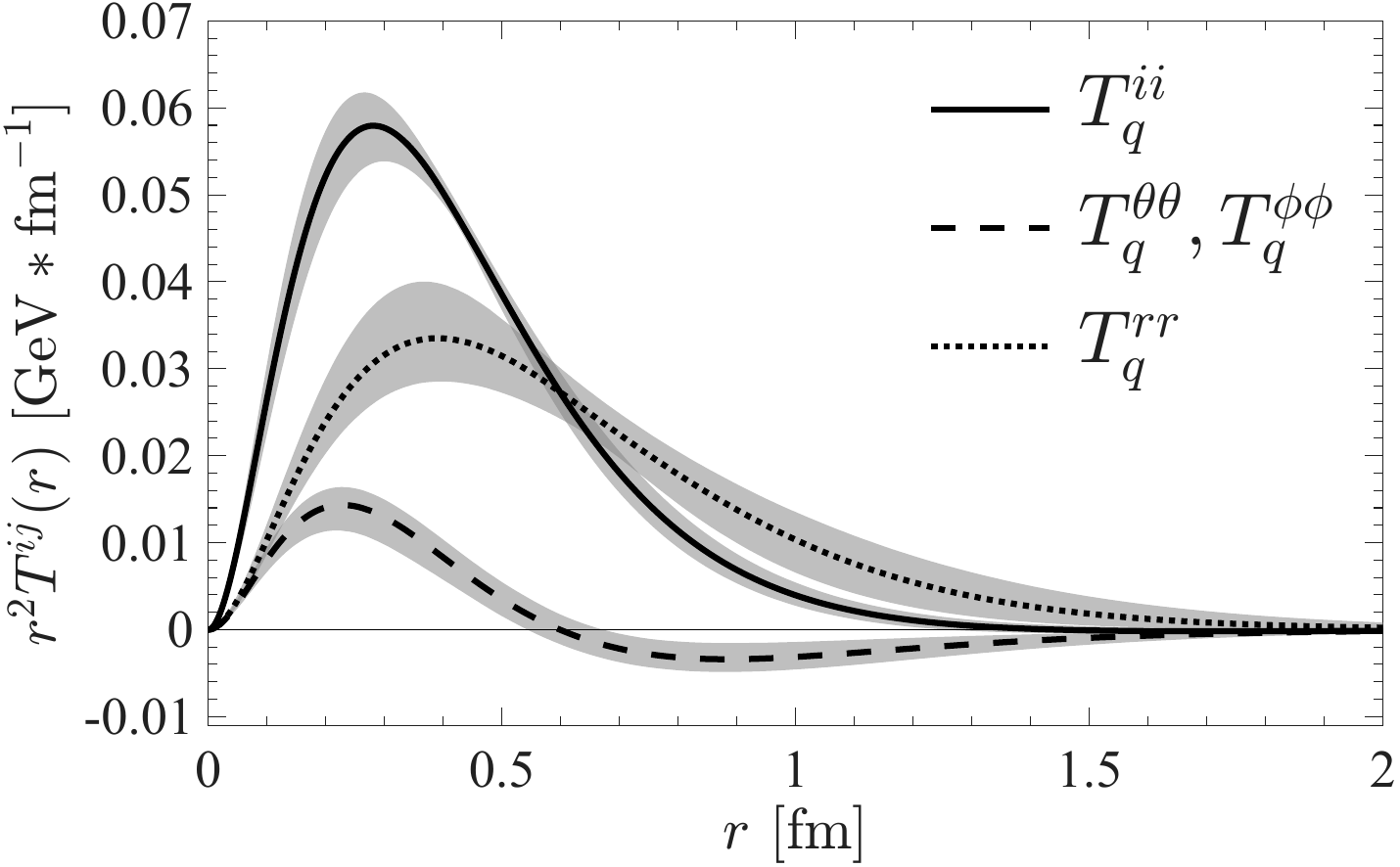}
    \caption{Quark kinetic MCDs in the proton, illustrated using the latest phenomenological fits to lattice QCD calculations and experimental data~\cite{Guo:2025jiz}: radial (dotted) and angular (dashed) MCDs, with the trace given as a sum (solid). The errors are estimated using MCMC samples with 90\% confidence intervals~\cite{Guo:2025jiz}, displayed as light-gray shaded areas. }
    \label{fig:Quark_Flow}
\end{figure}  

The quark kinetic MCD is shown in FIG.~\ref{fig:Quark_Flow} with radial (dotted line) and angular (dashed line) components in spherical coordinate. Different from the pure circular motions of the electron in the H-atom, quarks exhibit a combination of both radial and angular flow within the nucleon, analogous to an elliptical orbit in classical mechanics. The divergence-free contribution also indicates that this kinetic flow is sheared. However, as we previously discussed, this ordered momentum flow and its trace represent the kinetic energy of quarks, but is not readily related to the pressure/force due to the lack of discontinuities. 

\begin{figure}[ht]
    \centering
    \includegraphics[width=0.65\linewidth]{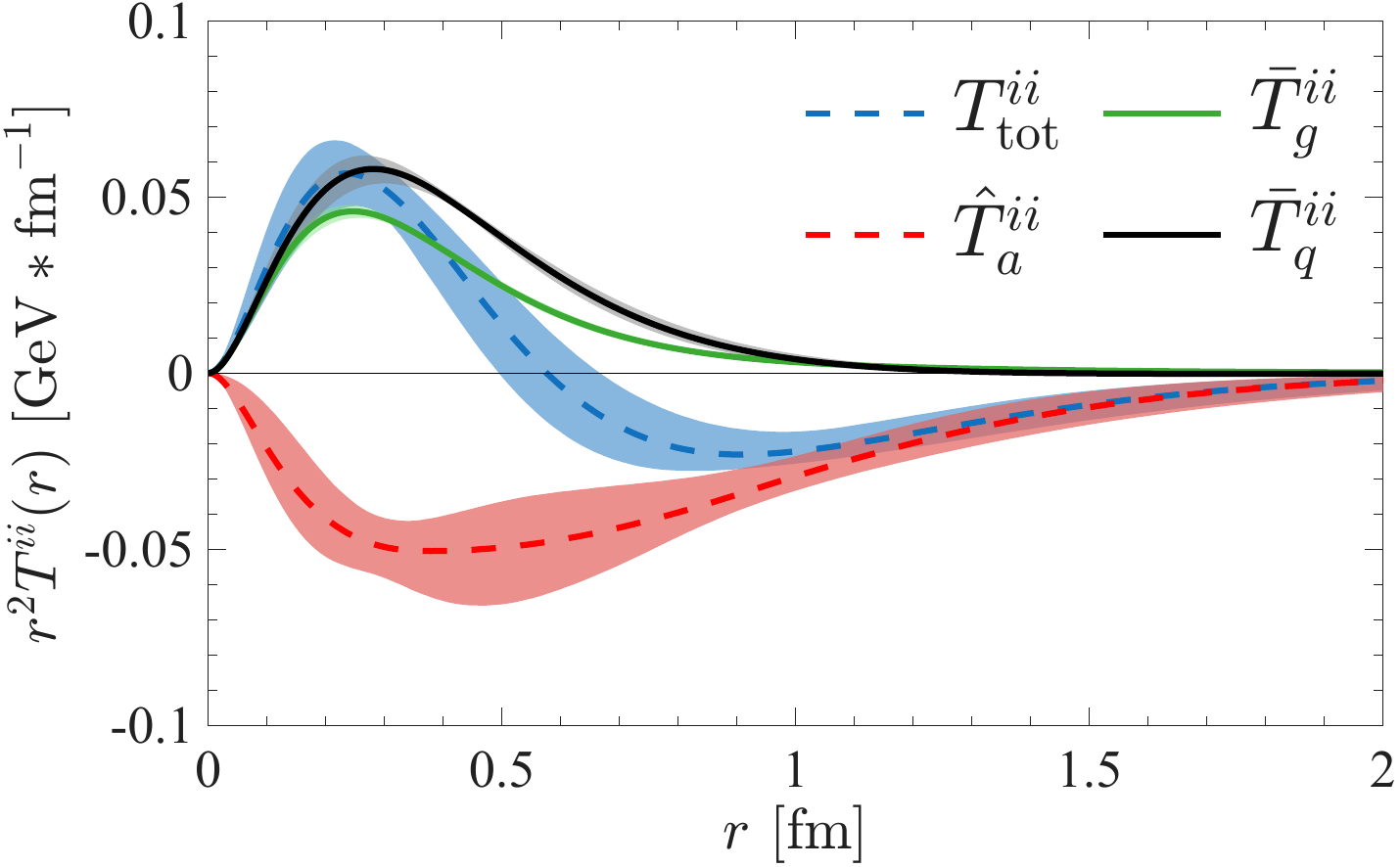}
    \caption{Trace of the momentum current distribution in proton and its decomposition into three components, demonstrated with the state-of-the-art results from lattice QCD calculations and experimental data~\cite{Guo:2025jiz}. The total current (blue dashed) satisfying the virial theorem in Eq.~(\ref{eq:von-Laue}) is decomposed into the positive quark kinetic (black solid) and gluon tensor (green solid) contributions and the negative trace anomaly (red dashed) contribution. Errors of 90\% confidence intervals are shown as shaded areas.}
    \label{fig:Proton_Trace}
\end{figure} 

We also plot the traces of above MCDs in FIG.~\ref{fig:Proton_Trace}. This plots differs a bit from that in Ref.~\cite{Ji:2025gsq} since here we have included a small contribution from the sigma form factor $F_\sigma$. This, however, does not change our discussions and conclusions in Ref.~\cite{Ji:2025gsq}. 
To be specific, though the Coulomb gluons and radiative gluons in the gluon tensor MCD $\langle\bar{T}^{ij}_g\rangle$ cannot be separated in a gauge-invariant and frame-independent way, its positive definiteness indicates that the radiative gluons may dominate the contribution, as opposed to the H-atom case, where the interference MCD is negative due to the dominance of Coulomb interaction~\cite{Ji:2022exr}. 

More interestingly, since the trace anomaly part has no corresponding momentum density, it reflects a pure interaction effect, which has been interpreted as the ``vacuum pressure''~\cite{Ji:1995sv, Liu:2023cse,Fujii:2025aip}. As we have seen before, in the M.I.T. bag model for the nucleon, the boundary separates the perturbative and non-perturbative vacuum and creates a discontinuity in the MCD. This discontinuity indicates a vacuum pressure/surface force, known as the cosmological constant. However, such a discontinuity does not exist in QCD, and it is unclear at this point what the pressure is acting on without a specific context. Therefore, this negative ``pressure'' $p_a(r)$ only indicates that quarks ``sweep'' out the true QCD vacuum and lower the expectation of gluon condensate, as also indicated in the instanton liquid model~\cite{Zahed:2021fxk}, consistent with the bag model phenomenology. Recent model calculations have also highlighted confining effects associated with the negative trace anomaly contribution to the MCD~\cite{Fujii:2025tpk,Tanaka:2025pny,Fujii:2025paw}. 

The quark and gluon contributions are found to dominate the small $r$ region of the total MCD, while the negative trace anomaly at large $r$ accounts for the total MCD being negative. This has depicted a picture for the structure of the proton as a hard core of quarks and gluons surrounded by the negative scalar (trace anomaly) cloud. It's also worth noting that this trace only represents the momentum flow rather than the pressure distribution within hadrons, and the von Laue condition for the total trace $\langle T^{ii}\rangle$ shown as the blue dashed line tells us nothing more than the momentum conservation. 

Note that the current data on the GFFs and MCDs show relatively large uncertainties on the trace anomaly and the total MCDs. This is because the large errors in determining the $C_{q,g}$ form factor, as show in FIG. 3 of Ref.~\cite{Guo:2025jiz}. Therefore, enhancing data precision and reducing errors are crucial in the future experimental and theoretical works.

\subsection{Seeking Forces on Quarks in the Nucleon}

For non-relativistic systems discussed in Sec.~\ref{sec:MCD-Classical}, force is a primordial concept, responsible for the total conservation and local changes of the momentum, and defines the interaction MCD. However, in quantum field theories such as QCD, local interactions among fields are more fundamental, leading to the MCD through Noether's theorem. While at high-energy where particles and their interactions can be described by perturbation theory, the quark and gluon fields are more natural to describe the non-perturbative structure of the nucleon as in lattice QCD calculations. Finding forces in the nucleon through interacting fields is a challenging and not necessarily unique exercise. 

Following the classical examples, we can define the force density on quarks through the divergence of the quark kinetic MCD, 
\begin{equation}
    \mathcal{F}^j \equiv \partial_i T^{ij}_{K} = \partial_i T^{ij}_q\ .
\end{equation}
In the static systems such as the nucleon, since the time-derivative term vanishes, we can make the above definition Lorentz covariant, 
\begin{equation}
    \mathcal{F}^j = \partial_\mu T^{\mu j}_{K} \ .
\end{equation}
After using the QCD equations of motion, one has~\cite{Braun:2004vf,Tanaka:2018wea},
\begin{equation}
    \mathcal{F}^j = g\bar{\psi}\gamma_{\mu}F^{\mu j}\psi = g\rho_{a}E_{a}^{j}+g(\vec{j}_{a}\times\vec{B}_{a})^{j}\ , 
\end{equation}
which is exactly the color-Lorentz force acting on quarks. However, the above equation is only formal because the operator on the right-hand side has power divergences (twist-4) and mixes with the vector current of quarks $\bar \psi \gamma^j \psi$. Therefore, one needs additional renormalization subtractions that render the operator physically more ambiguous. A variant of the above operator $g\bar{\psi}\gamma^{(\mu}F^{\nu) j}\psi$ with $\mu\nu$ symmetric, a twist-3 operator with only logarithmic divergences, has been considered as a color-Lorentz force on quarks in deeply inelastic processes and studied in lattice QCD calculations and instanton liquid model~\cite{Filippone:2001ux,Burkardt:2008ps,Abdallah:2016xfk,SANE:2018pwx,Aslan:2019jis,Burger:2021knd,Crawford:2024wzx,Liu:2025ypg}. 

\begin{figure}[ht]
    \centering
    \includegraphics[width=0.65\linewidth]{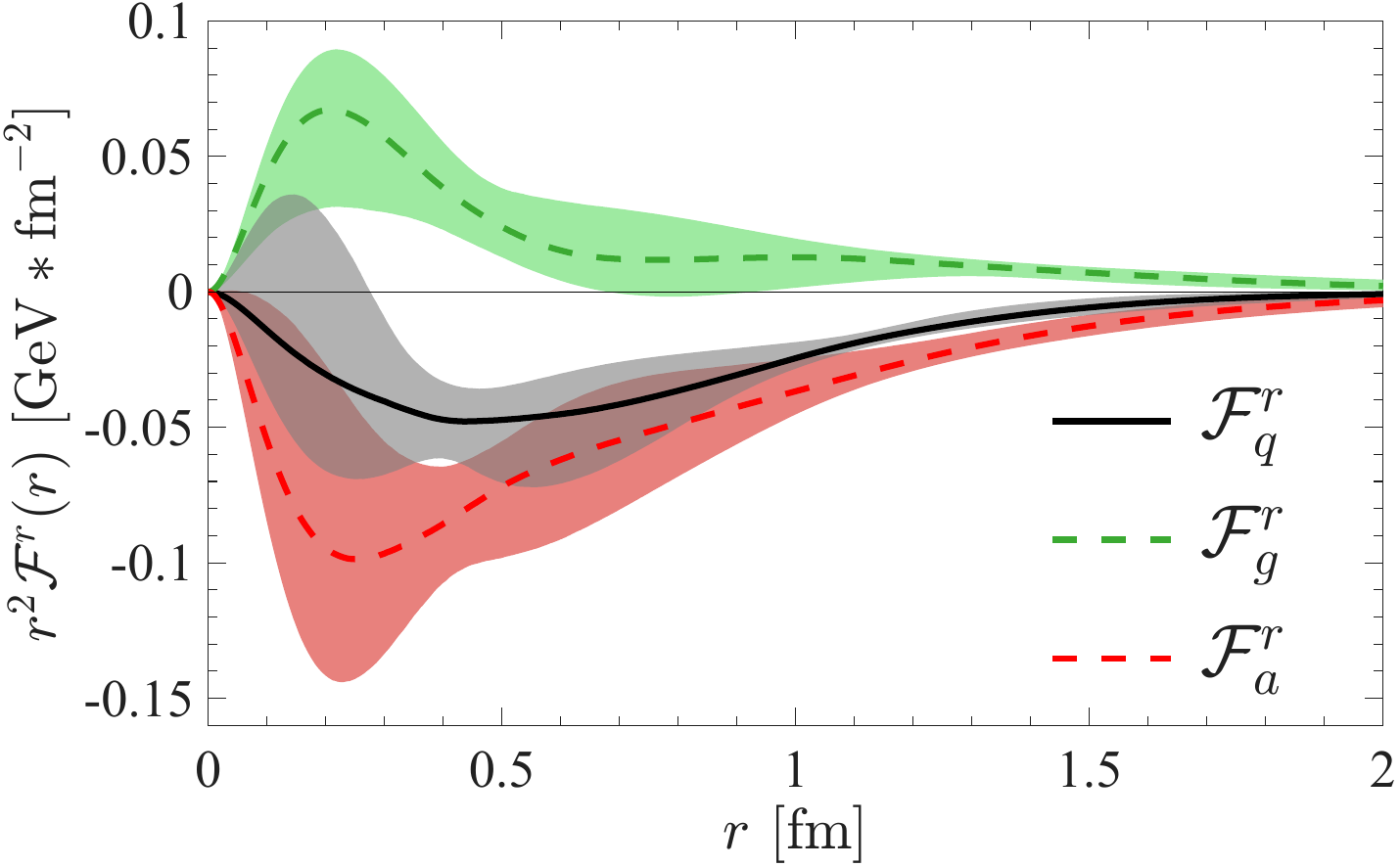}
    \caption{Radial color-Lorentz force density distributions acting on quarks in the proton, analyzed using the latest results from lattice QCD calculations and experimental data~\cite{Guo:2025jiz}: the large attractive force from the anomaly (red dashed) and the repulsive force from the gluon tensor (green dashed) combine to yield the overall confining force (black solid) on quarks. Errors of 90\% confidence intervals are shown as shaded areas.}
    \label{fig:Forces}
\end{figure}

Here we choose a definition of the force on quarks by studying the divergences of the {\it renormalized MCD matrix elements}. To be specific, we advocate a definition~\cite{Polyakov:2018exb,Won:2023zmf},  
\begin{equation}
  {\cal F}_q^j (\vec{r}) \equiv \partial_i \langle T^{ij}_q \rangle(\vec{r}) 
  = \frac{M}{4}\nabla^j \left[ G_{s,q}(r) - F_\sigma(r) \right]\ . 
\end{equation}
This color-Lorentz force can be also understood as the central one acting on the quark due to its kinetic motion. On the other hand, momentum continuity equation allows us to trace the force to the divergences of the gluon tensor and anomaly MCDs,
\begin{equation}
    \partial_i \langle T^{ij}_q \rangle (\vec{r})=- \partial_i \langle T^{ij}_g \rangle(\vec{r}) - \nabla^j p_a(\vec{r}) \equiv {\cal F}_{g}^j(\vec{r}) + {\cal F}_{a}^j(\vec{r}) \ ,
\end{equation}
where these two contributions can be further written as, 
\begin{align}
    \mathcal{F}_{g}^{j}&=-\partial_{i} \langle T_{g}^{ij} \rangle(\vec{r})=-\frac{M}{4}\nabla^{j}G_{s,g}(r)\\
    \mathcal{F}_{a}^{j}&=-\partial_{i} \langle T_{a}^{ij} \rangle(\vec{r})=+\frac{M}{4}\nabla^{j}\left[G_{s}(r)-F_{\sigma}(r)\right] \equiv -\nabla^j p_a
\end{align}
These forces are exclusively along the radial direction. Using the previously mentioned parametrization and parameters extracted from lattice and experimental data, we plot the radial force densities on quarks from the separate gluon tensor and anomaly contributions in FIG.~\ref{fig:Forces}. Again, this result only slightly differs from that in Ref.~\cite{Ji:2025gsq} by a small contribution from the sigma term $m\bar{\psi}\psi$. From the plot, one finds that the gluon tensor imposes a repulsive force, indicating perhaps the dominance of radiative gluons. More interestingly, the trace anomaly exerts a larger negative binding force on quarks.

With the above numerical results, we can calculate the average force on quarks from the trace anomaly by integrating the force density over $\vec{r}$, 
\begin{equation}
    \int {\rm d}^3 \vec{r} ~{\cal F}_a(\vec{r}) = -0.99^{+0.12}_{-0.11} ~{\rm GeV/fm} \ .
\end{equation}
One finds that the average force from gluon scalar (trace anomaly) has a magnitude similar to the confinement string tension~\cite{Sun:2020pda,Brambilla:2021wqs}, roughly 1 GeV/fm, a concrete evidence that the anomaly plays an central role in the confinement effect~\cite{Rothe:1995hu,Liu:2021gco,Shuryak:2021yif,Liu:2024rdm}. Meanwhile, one can also define a radius of the force on quarks as,
\begin{equation}
    \left\langle r^2 \right\rangle_{{\cal F}_a} = \frac{\int {\rm d}^3 \vec{r} ~ r^2 \mathcal{F}_a(\vec{r})}{\int {\rm d}^3 \vec{r} ~ \mathcal{F}_a(\vec{r})} = \left(0.80^{+0.08}_{-0.07}\ {\rm fm}\right)^{2}
\end{equation}
This radius will also provide an important information for the validity of these local densities in the Breit frame, as previously discussed.

\section{Comments on Polyakov's Interpretation} \label{sec:Argument}

The mechanical effect encoded in a nucleon's GFFs was proposed by Polyakov and collaborators, stating that the $C/D$-form factor characterizes the spatial distributions of pressure and shear forces~\cite{Polyakov:2002wz,Polyakov:2002yz}. In their view, the nucleon approximates a mechanical continuum system (like a fluid), and the total static MCD inside a hadron with spherical symmetry is decomposed into the 3D trace and traceless parts, 
\begin{equation}
    T^{ij}(\vec{r})= p(r)\delta^{ij} + s(r)\left(\frac{r^{i}r^{j}}{r^{2}}-\frac{1}{3}\delta^{ij}\right) \ ,
    \label{eq:polyakov}
\end{equation}
where the monopole $p(r)$ and tensor $s(r)$ terms are interpreted, respectively, as the ``pressure'' and ``shear'' forces. They further emphasize their meanings as the contact forces within the nucleon experienced by different ``parts''. That is, the momentum flow through a surface element, $T^{ij}{\rm d}S^j$, is now a force acting on it from the one side of the surface to the other. Under such a physical picture, the ``mechanical stability'' of a nucleon was then proposed in~\cite{Perevalova:2016dln}. 

In this section, we comment on the above interpretations from what we have learned in Secs.~\ref{sec:MCD-Classical} and \ref{sec:MCD-Quantum}. We conclude that there is little justification for Polyakov's mechanical interpretation in the sense that $p(r)$ and $s(r)$ have no clear physical significance. The $C/D$ form factor only depicts one overall pattern of the momentum flow with built-in mechanical stability. 

\subsection{Kinetic MCD and Anisotropic Flow}

In many examples we have discussed in Secs.~\ref{sec:MCD-Classical} and \ref{sec:MCD-Quantum}, kinetic MCD plays an important role in the momentum flow. This is physically the clearest and most unambiguous momentum transport mechanism. In classical systems such as gases and liquids as well as the quantum Fermi gas, the kinetic MCD is isotropic and gives rise to a contribution of kinetic pressure $p_K$. However, when there is an overall macroscopic flow of particles, the kinetic MCD is anisotropic. For example, in the case of fluids, there is a contribution $\rho v^iv^j$.  In the ground state of H-atom or many other quantum systems, the kinetic MCD is also anisotropic. In these cases, $T^{ii}_K/3$ cannot simply be interpreted as a pressure.
Furthermore, the traceless part of $\rho v^i v^j$ simply reflects the anisotropy of the momentum flow. It has little to do with the shear force of twisted deformation in solids or the shear stress due to dissipative momentum transfer in liquids. 

The nucleon MCD has kinetic contribution as a part of its total, $T^{ij}_{q+g}$ (note that the gluon part here is not entirely from the physical gluon quanta, but also receives contributions from Coulomb gluons which we will ignore in the discussion here.) In the nucleon interior,  quarks are moving relativistically to generate, among others, the nucleon's magnetic moment, and a part of the nucleon mass comes from the kinetic energies of quarks and gluons. However, the kinetic part of the MCD is anisotropic, and according to the above, the quark and gluon MCD monopole term $T^{ii}_{q+g}/3$ cannot be simply interpreted as the kinetic pressure, let along a normal force between adjacent parts of the nucleon. The 3D traceless or the tensor part is also unrelated to shear forces; it is simply reflecting an ordered flow of momentum.

\subsection{Interaction MCD and Long-Range Force}
 
Is it possible then to interpret the interaction part of the nucleon MCD as pressure and shear forces? The answer lies in whether the interaction in the nucleon can be considered
as short-ranged. In the case of a large nucleus where the range of nucleon-nucleon interaction is comparable to the size of the nucleon, as discussed in Sec.~\ref{sec:nuclmattter}, the interaction MCD is isotropic and can be considered as surface forces between different parts of the nucleus. However, the QCD color-Lorentz forces inside the nucleon have a range comparable to its size, namely about 1~fm, which are not screened. This situation is similar to the static Coulomb interactions in the Thomas-Fermi model of atoms in Sec.~\ref{sec:fermi}, and the respective static Coulomb contribution to the MCD cannot be described by an interaction pressure alone. Therefore, the static color-force contribution in MCD $T^{ij}_g$ has no simple mechanical significance other than that its divergence generates a force density. It has little to do with shear surface forces as in solids or fluids because of the long-range nature. As we discussed in the previous section, the trace anomaly contribution is proportional to $\delta^{ij}$ and can be regarded as a pressure. But it is not directly a surface because it has no discontinuity. Rather, it serves as a potential whose divergence generates a force density. 

\subsection{$C/D$ Form Factor For Overall Momentum Flow}

Fundamentally, MCD arises from describing the local conservation of momentum. Therefore, different parts of MCD provide different mechanisms for the momentum transport to happen. The non-diagonal part of $T^{ij}$ represent the shear flow of momentum, i.e., momentum in the $i$-th direction is flowed in the $j$-th direction. 

When the trace $p(r)$ is positive, the momentum flows in the positive direction, and a negative $p(r)$ means that it flows in the negative direction. And the momentum conservation requires that the positive and negative flows are balanced out. For any bound states, this condition is satisfied. 

In this overall pattern of momentum flow, there is no net force
involved. The net force density is zero everywhere and there are no repulsions and attractions. The non-diagonal part of $T^{ij}$ does not give rise
to any shear forces. 

There is an infinite number of such local momentum balance pictures. 
The one couples to gravity is interesting because it can in principle be measured through a different interaction. It is also interesting that
it can be probed through the high-energy scattering. This particular momentum flow is also related to the mass and possibly scalar confinement structure of the nucleon~\cite{Ji:2021mtz}.

\subsection{Mechanical Stability Conditions}

Classical mechanics was abandoned for quantum mechanics because the former cannot explain the stability of atoms. In quantum mechanics, the stability of a system takes a different form: so long as the Hamiltonian of the system is hermitian, its ground state must be stable because there is no lower energy state it can decay to. When applying this argument to the nucleon, there is no doubt that the QCD will make it a stable system because it is the lowest energy state with the conserved baryon number 1. Therefore, it is perplexing to use the stability condition of a classical system to argue the stability of a QCD nucleon. 

The first classical ``mechanical stability condition'' proposed in~\cite{Laue:1911lrk} has been called the von Laue condition in the literature. Historically, this condition (Eq.~(\ref{eq:von-Laue})) might have played an important role in finding a complete theory for certain physics. However, in a stationary state of a well-defined system, it is an automatic consequence of momentum conservation, independent of whatever system under consideration.  
With the relation $\langle T^{ii}\rangle(\vec{r}) =\varepsilon_q(\vec{r})+\varepsilon_g(\vec{r})-3{\varepsilon}_a(\vec{r})$ in the nucleon, this directly gives the virial theorem between the scalar and tensor energy contributions to the nucleon as $M_T = 3M_S$~\cite{Ji:1994av,Ji:1995sv,Lorce:2025oot}.  This constraint also leads to the familiar virial theorem for non-relativistic systems. 

The von Laue condition implies that there must exist at least one node for the momentum flow $p=\langle T^{ii}\rangle/3$. Following this, it has been proposed that system is stable if one has the ``repulsive forces'' in the internal region $p(r)>0$ are balanced by the ``attractive forces'' in the outer area $p(r)<0$, (see e.g. Ref.~\cite{Goeke:2007fp,Goeke:2007fq,Cebulla:2007ei,Burkert:2018bqq,Polyakov:2018zvc}), which intuitively leads to a negative $D$-term. However, there are many counter examples in which the system has positive $p(r)$ at large $r$ and negative inside~\cite{Ji:2022exr,Fu:2023ijy} and positive $D$-term~\cite{Ji:2022exr,Fu:2022rkn,Fu:2023ijy,Wang:2023bjp}. Therefore, further requiring the sign of $p(r)$ for stability argument cannot be rigorously established.

\section{Conclusions}\label{sec:Conclusion}

In this paper, we have examined the physics of the momentum current density and associated mechanical effects in various physical systems. We have also studied the physics of several contributions to the total MCD of the system. The MCD receives contributions from both kinetic motion of particles and the interacting fields within. While the kinetic MCD can be unambiguously defined as flow of particles carrying momenta, interaction MCD acquires extra gauge degrees of freedom, called the superpotential. Though there is a preferable version that is believed to be the gravitational charge, it appears to have no mechanical advantages over the others. 

Despite the non-uniqueness of the MCD, the mechanical effects can be unambiguously determined through the divergence of the kinetic MCD, interpreted as the force density exerted on particles. Interestingly, the discontinuities of the MCD directly corresponds to a well-defined surface force, such as the thermal pressure of ideal gases on the boundary. Meanwhile, in the case of contact interactions such as in solids and liquids, a certain choice of interaction MCD, the stress tensor, represents the surface forces including pressures and shears as well. 

For the MCD in the nucleon, because of the anisotropic flow of momentum, contribution of the static color fields, and the long-range nature of the color forces, the interpretation in Eq.~(\ref{eq:polyakov}) suggested by Polyakov et al has no clear physics basis. 

Finally, we give an extended discussion about the QCD MCD in the nucleon based on our previous publication~\cite{Ji:2025gsq}. It was found that the scalar form factor $G_s(q^2)$ encodes the primary information about the mechanical forces. Quarks are found to experience a strongly confining color-Lorentz force from the gluon tensor and gluon scalar (trace anomaly) contributions. While the gluon tensor contributes a repulsive force, indicating the radiation dominance, the trace anomaly provides a large attractive force with average magnitude around 1 GeV/fm, which is similar to the well-known QCD string tension. 

\acknowledgments

We thank Adam Freese, Daisuke Fujii, Victor Galitski, Mamiya Kawaguchi, Keh-Fei Liu, Zein-Eddine Meziani, Gerald A. Miller, Peter Schweitzer, Yushan Su, Mitsuru Tanaka, Jinghong Yang, Feng Yuan, and Ismail Zahed for useful discussions. We also thank Yuxun Guo for providing the data for the parametrization of GFFs. X. Ji and C. Yang are partially supported by Maryland Center for Fundamental Physics (MCFP). C. Yang also acknowledges partial support by the U.S. Department of Energy, Office of Science, Office of Nuclear Physics under the umbrella of the Quark-Gluon Tomography (QGT) Topical Collaboration with Award DE-SC0023646.

\appendix
\section{Total Derivative in Lagrangian and the Equation of Motion}\label{sec-app:EoM}
The total derivative of the Lagrangian can be written as,
\begin{equation}
    \delta\mathcal{L}=\sum_{n=0}^{\infty}\frac{\partial S^{\rho}}{\partial\left(\partial_{\nu_{1}}\dots\partial_{\nu_{n}}\phi\right)}\left(\partial_{\nu_{1}}\dots\partial_{\nu_{n}}\right)\left(\partial_{\rho}\phi\right)\equiv\sum_{n=0}^{\infty}D^{\nu_{1}\dots\nu_{n}}S^{\rho}\left(\partial_{\nu_{1}\dots\nu_{n}\rho}\phi\right)
\end{equation}
where we adopt the following notation for simplicity, 
\begin{equation}
    \partial_{\mu_{1}\dots\mu_{n}} \equiv \partial_{\mu_{1}}\dots\partial_{\mu_{n}}; ~~~ D^{\nu_{1}\dots\nu_{n}}\equiv\frac{\partial}{\partial\left(\partial_{\nu_{1}\dots\nu_{n}}\phi\right)}
\end{equation}
Therefore, the additional part of the equation of motion can be written as, 
\begin{equation}
    \delta {\rm EoM} \equiv \sum_{n=0}^{\infty}\left(-1\right)^{n}\left(\partial_{\mu_{1}\dots\mu_{n}}\right)D^{\mu_{1}\dots\mu_{n}}\delta\mathcal{L} \ , 
\end{equation}
which can be further simplified as follows, 
\begin{align}
    D^{\mu_{1}\dots\mu_{n}}\delta\mathcal{L}
    =&D^{\mu_{1}\dots\mu_{n}}\left[\sum_{m=0}^{\infty}D^{\nu_{1}\dots\nu_{m}} S^{\rho}\left(\partial_{\nu_{1}\dots\nu_{m}\rho}\phi\right)\right]\\
    =&\sum_{m=0}^{\infty}\left[D^{\mu_{1}\dots\mu_{n}}D^{\nu_{1}\dots\nu_{m}}S^{\rho}\left(\partial_{\nu_{1}\dots\nu_{m}\rho}\phi\right)+\left(D^{\nu_{1}\dots\nu_{m}}S^{\rho}\right)\frac{\partial\left(\partial_{\nu_{1}\dots\nu_{m}\rho}\phi\right)}{\partial\left(\partial_{\mu_{1}\dots\mu_{n}}\phi\right)}\right]\\
    =&\partial_{\rho}\left[D^{\mu_{1}\dots\mu_{n}}S^{\rho}\right]+\sum_{m=0}^{\infty}\left(D^{\nu_{1}\dots\nu_{m}}S^{\rho}\right)\delta_{\nu_{1}}^{(\mu_{1}}\dots\delta_{\nu_{m}}^{\mu_{n-1}}\delta_{\rho}^{\mu_{n})} \delta_{m+1,n}
\end{align}
Now one finds, 
\begin{equation}
    \delta {\rm EoM} = \sum_{n=0}^{\infty}\left[\left(-1\right)^{n}+\left(-1\right)^{n+1}\right]\left(\partial_{\mu_{1}\dots\mu_{n}}\right)\left[\partial_{\rho}\frac{\partial S^{\rho}}{\partial\left(\partial_{\mu_{1}\dots\mu_{n}}\phi\right)}\right]=0 \ .
\end{equation}
The additional part exactly vanishes, indicating that the total derivative in the Lagrangian will not change the equation of motion. This can also be seen from the integral form of the principle of least action, where the total derivative term vanishes as a boundary term. 

\section{Constraining the Divergence in Self Energy of a Point-like Charge}\label{sec-app:point_like_charge}

In the case of H-atom, the MCD for the self electric field of the electron is given by, 
\begin{equation}
    \langle T_{e}^{ij}(\vec{r})\rangle=\frac{\hbar^{2}}{4\pi^{2}m_{e}a_{0}^{4}}\int{\rm d}^{3}r_{1}e^{-2r_{1}/a}\frac{1}{\left|\vec{r}_{1}-\vec{r}\right|^{4}}\left[\frac{1}{2}\delta^{ij}-\frac{\left(\vec{r}-\vec{r}_{1}\right)^{i}\left(\vec{r}-\vec{r}_{1}\right)^{j}}{\left|\vec{r}_{1}-\vec{r}\right|^{2}}\right] \ .   
\end{equation}
Although this MCD is conserved itself, it is divergent with a positive-definite trace, $T_e^{ii}>0$, which invalidates the von-Laue condition as in Eq.~(\ref{eq:von-Laue}). This is due to the singularity near the point-like particle. This can be resolved by using the following relation, 
\begin{equation}
     \left(\delta^{ij}\nabla^{2}-\nabla^{i}\nabla^{j}\right)\frac{1}{r^{2}}=\frac{8}{r^{4}}\left(\frac{1}{2}\delta^{ij}-\frac{r^{i}r^{j}}{r^{2}}\right) \ ,
\end{equation}
and taking the derivative out the integral, and one obtains a finite MCD, 
\begin{equation}
    \langle T_{e}^{ij}(\vec{r})\rangle=\left(\delta^{ij}\nabla^{2}-\nabla^{i}\nabla^{j}\right)\left[\frac{\hbar^{2}}{32\pi^{2}ma_{0}^{4}}\int{\rm d}^{3}r_{1}e^{-2r_{1}/a}\frac{1}{\left|\vec{r}-\vec{r}_{1}\right|^{2}}\right]
\end{equation}
It is easy to check that the integral is finite and the above MCD now satisfies the von-Laue condition. This process effectively introduces a UV regulator to the MCD, which creates a large negative region near the point-like particle and thereby eliminates the divergence and restores the von-Laue condition. The full relation with explicit UV regulator is, 
\begin{equation}
    \left(\delta^{ij}\nabla^{2}-\nabla^{i}\nabla^{j}\right)\frac{1}{r^{2}}=\frac{8}{r^{4}}\left(\frac{1}{2}\delta^{ij}-\frac{r^{i}r^{j}}{r^{2}}\right)\theta(\vec{r} \notin u)-\frac{\delta^{ij}}{3}\delta^{(3)}(\vec{r})\int_{\vec{r}^{\prime}\notin u}{\rm d}V^{\prime}\frac{4}{(r^{\prime})^{4}}
\end{equation}
In obtaining the electric MCD of the electron, one performs an average over the wave function, therefore this large negative small region is now distributed over space, leading to a finite result. Similarly, one can write the electric MCD of the proton as, 
\begin{align}
     T_{p}^{ij}(\vec{r})
     =\frac{\hbar^{2}}{4\pi m_{e}r^{4}a_{0}}\left(\frac{1}{2}\delta^{ij}-\frac{r^{i}r^{j}}{r^{2}}\right)=\left(\delta^{ij}\nabla^{2}-\nabla^{i}\nabla^{j}\right)\left[\frac{\hbar^{2}}{32\pi ma_{0}r^{2}}\right] \ ,
\end{align}
which is also conserved by itself. 

\section{Decomposition of QCD EMT and Ambiguities on $\bar{C}$ Form Factor}
In the originally proposed parametrization of the QCD EMT using GFFs~\citep{Ji:1996ek}, a $\bar{C}$ form factor is introduced to account for the non-conservation terms, 
\begin{align}
    \left\langle P^{\prime}\left|T_{i}^{\mu\nu}\right|P\right\rangle =
    \bar{U}&\left(P^{\prime}\right)\left[A_{i}(q^2)\gamma^{(\mu}\bar{P}^{\nu)}+B_{i}(q^2)\frac{\bar{P}^{(\mu}i\sigma^{\nu)\alpha}q_{\alpha}}{2M}\right. \nonumber \\
    &  \left.+C_{i}(q^2)\frac{q^{\mu}q^{\nu}-q^{2}g^{\mu\nu}}{M}+\bar{C}_{i}(q^2)Mg^{\mu\nu}\right]U\left(P\right) \ ,
\end{align}
where $i=q,g$ represents quarks and gluons. However, there are different decompositions of the EMT operator, and $\bar{C}_i$ is closely related to these definitions. In Ji's three-term decomposition~\cite{Ji:1994av,Ji:1995sv}, the quark and gluon tensor EMTs are required to be traceless. Therefore, $\bar{C}_{q,g}$ becomes the quark and gluon scalar form factors as shown in Eq.~(3) and (4) of~\cite{Ji:2025gsq}. 

More generally, there could be other options for decomposing the EMT operators. For example, the decomposition in the present paper is based on Ji's decomposition with an extra rearrangement, as shown in Eq.~(\ref{eq:EMT-QCD}). Consequently, the trace of quark EMT is required to be the sigma term $\bar{\psi}m\psi$, while the gluon tensor EMT is traceless. Therefore, $\bar{C}_{q,g,a}$ is expressed using the quark and gluon scalar form factors and the sigma term form factor, as shown in Eq.~(\ref{eq:FF-q}) - (\ref{eq:FF-tr}), which can be written as, 
\begin{align}
    \bar{C}_{q}(q^2)&=-\frac{1}{4}\left[G_{s,q}(q^2)-F_{\sigma}(q^2)\right]\\
    \bar{C}_{g}(q^2)&=-\frac{1}{4}G_{s,g}(q^2)\\
    \bar{C}_{a}(q^2)&=+\frac{1}{4}\left[G_{s,q}(q^2)+G_{s,g}(q^2)-F_{\sigma}(q^2)\right]
\end{align}
In summary, $\bar{C}$ depends on the choices of EMT operators. Only when a specific version of EMT is given, can its expression be derived unambiguously.


\end{document}